\documentclass{article}
\usepackage{preamble}

\commentfalse
\commenttrue

\title{Compiling Low-Magic Protocols to have Short Messages}

\author{Owen Boseley, Darin Ershov, Anthony Mangus, Chinmay Nirkhe, and Toby Thornburg}
\affil{University of Washington}

\date{}
\begin{document}

\maketitle
\begin{abstract}
We develop new circuit compilation techniques to reduce communication complexity in various quantum verification tasks. We show that magic complexity---the number of non-Clifford gates in a quantum circuit---provides a good upper bound on the total communication required for $\QMA[k]$. We then provide evidence that the same bound holds for $\QIP$ by proving $\QMAM$ has similar communication compilations and non-adaptive Clifford $\QIP$ is polynomial-time solvable. In the process, we develop new techniques to analyze the entanglement structure of the stabilizer projectors associated with the verifier, and mathematics to represent Clifford interactive protocols as a product of stabilizer projectors.
\end{abstract}
\section{Introduction}
Quantum verification protocols provide a general framework for studying tasks achievable by a computationally limited client, known as Arthur, and an untrusted but powerful quantum server, known as Merlin. A poignant example is quantum delegation, where Arthur uses Merlin to perform a significantly more complex computation while verifying Merlin's work to maintain a high likelihood of correctness. Three natural variants of these communication protocols are: $\QMA$, the quantum version of $\NP$; $\QMA[k]$, where Arthur receives unentangled messages from $k$ Merlins; and $\QIP$, a multi-round protocol where the parties alternate exchanging messages. In each of these, communication length serves as a good primitive for understanding hardness; higher communication protocols enable Arthur to solve harder tasks.

We investigate the power of these verification procedures through the lens of computation restriction. One well-studied notion of computation restriction is magic complexity, the number of non-Clifford gates used in a circuit. There are two reasons this is a useful measure. First, magic gates seem to capture the effective ``quantumness'' of a circuit. A purely Clifford circuit is known to be classically simulatable in polynomial time, while the best known bound for non-Clifford circuits scales exponentially in the magic complexity. Second, realizing magic gates is far more expensive and error-prone than Clifford gates. Thus, it is interesting to ask about the power of quantum computation even with limited magic complexity. We explore this question using circuit compilation: can low-magic protocols be recompiled to use low communication?

\subsection{The parameterized complexity of quantum verification}

The relationship between communication and magic in verification was first studied by Arunachalam, Bravyi, Nirkhe, and O'Gorman~\cite{ABNO2022}. They give an upper bound on the witness length for $\QMA$ protocols through circuit compilation: given a $\mathsf{QMA}$ verification circuit with $t$ $T$-gates and any number of Clifford gates, they showed how to find and efficiently describe, in classical polynomial time, a stabilizer subspace of dimension at most $2^t$ containing an optimal witness. Equivalently, Arthur can efficiently compile a Clifford precomputation gadget, ask Merlin for a shorter witness of length at most $t$ qubits, and reconstruct an optimal witness to the original protocol by adjoining ancilla and applying the gadget. The bound holds even when acceptance requires a prescribed measurement outcome on many output qubits. It also leads to a classical randomized algorithm for estimating the optimal acceptance probability whose running time, at fixed relative precision, is polynomial in the circuit size and singly exponential in $t$.

This result can be seen as a natural non-deterministic extension of the Gottesman-Knill principle~\cite{Gottesman1997}: Clifford and low-magic circuits on a stabilizer input admit efficient classical simulation. Arunachalam \emph{et. al.}~\cite{ABNO2022} extend this principle from simulation to verification, i.e., extending from the class $\BQP$ to the class $\QMA$. In verification, Merlin must \textit{find} an optimal input to Arthur's circuit, which could be arbitrarily (polynomially) large. Despite the unrestricted nature of the input, the precomputation gadget shows that Arthur's magic complexity bounds the search space that needs to be optimized over. Magic complexity, through this compilation technique, limits the communication complexity of the verification circuit, which in essence bounds the hardness of such problems that it can solve. %

However, this compilation technique does not directly apply to other verification complexity classes such as the larger classes of $\QMA[k]$ and $\QIP$. Each of these classes enforces a structural constraint on Merlin's witness: the $\QMA[k]$ witness must be $k$ unentangled states, and the $\QIP$ witness must be interleaved with Arthur's messages. The precomputation gadget $\QMA$ could violate these constraints, allowing Merlin to `trick' Arthur's verification. Therefore, it is not obvious \emph{a priori} that magic complexity should bound communication length beyond $\QMA$. In this paper, we show that this bound does indeed hold in $\QMA[k]$ and may extend to $\QIP$. To show this, we develop new compilation techniques that preserve the guarantees of each class.

\subsection{Results}
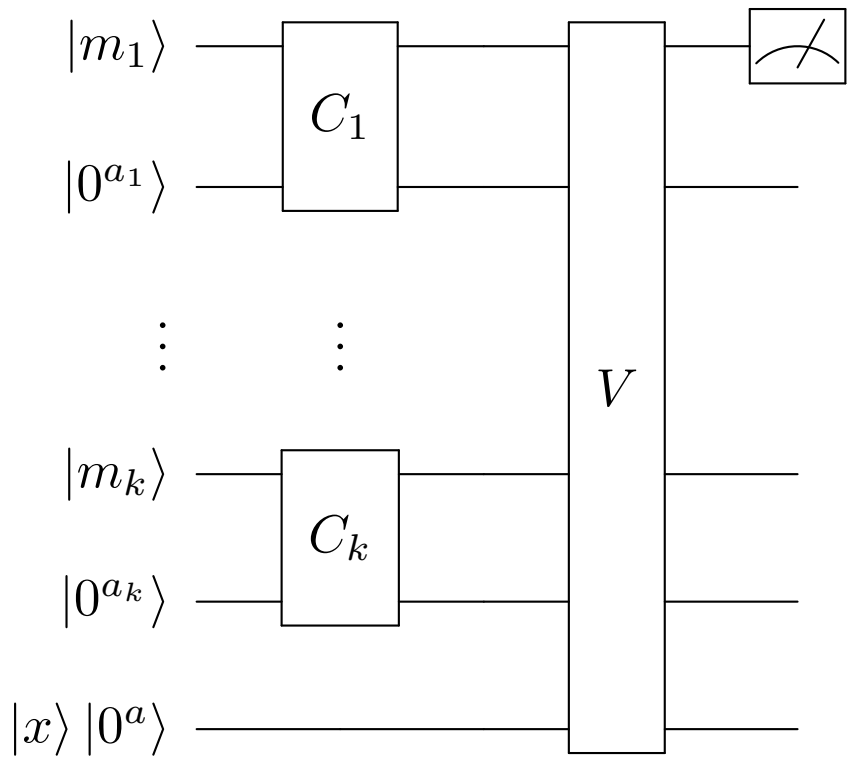
\begin{figure}[H]
    \centering
    \makebox[\textwidth][c]{%
        \Qcircuit @C=1.6em @R=1.5em {
        \lstick{\ket{m_1}}
            & \multigate{1}{C_1} & \qw
            & \multigate{5}{V} & \meter \\
        \lstick{\ket{0^{a_1}}}
            & \ghost{C_1} & \qw
            & \ghost{V} & \qw \\
        \lstick{\vdots}
            & \push{\vdots} &
            & \nghost{V} & \\
        \lstick{\ket{m_k}}
            & \multigate{1}{C_k} & \qw
            & \ghost{V} & \qw \\
        \lstick{\ket{0^{a_k}}}
            & \ghost{C_k} & \qw
            & \ghost{V} & \qw \\
        \lstick{\ket{x}\ket{0^a}}
            & \qw & \qw
            & \ghost{V} & \qw
        }
    }%
    \caption{The compiled $\QMA[k]$ verifier $V'$.}
    \label{fig:qmak_transformed}
\end{figure}

For a $\QMA[k]$ verifier circuit $V$ solving some problem on input $x$, let $p(V,x)$ be the maximum acceptance probability of $V$ that Merlin can achieve on $x$ over all legal proofs. We provide an algorithmic compilation procedure to produce a modified $V'$ with $p(V',x)=p(V,x)$. This $V'$ takes the form of $k$ efficiently computable gadgets, each applied to a small witness and fresh ancilla, followed by verification by $V$. %
This compilation is summarized via the following theorem and shown in Figure \ref{fig:qmak_transformed}.

\begin{theorem}[Informal] \label{thm:qmak}
Let $V$ be a $\QMA[k]$ verifier whose circuit contains $t$ $T$ gates and measures one output qubit. There is an efficiently computable circuit $V'$ such that each $V'$ accepts at most $t+1$ qubits per witness such that $p(V',x)=p(V,x)$. This circuit $V'$ applies Cliffords $C_j$ to witness $j$ padded with some number of $\ket0$, then verifies with $V$. In total, $V'$ has proof length at most $k(t+1)$.
\end{theorem}

Furthermore, one can extend this to show that if $V$ measures $o$ output qubits, then the resulting circuit $V'$ will accept witnesses of size $t+o$ qubits per witness and thus have total proof length $k(t+o)$.

For $\QMA[2]$ specifically, a different technique allows us to bound the proof length independent of the number of output qubits. This compilation also produces precomputation circuits that act locally to each of the two witnesses, as in Figure \ref{fig:qmak_transformed}.

\begin{theorem}[Informal] \label{thm:qma2}
Let $V$ be a $\QMA[2]$ verifier whose circuit contains $t$ $T$ gates and measures any number of output qubits. There exists a circuit $V'$ such that each $V'$ accepts at most $2t$ qubits per witness such that $p(V',x)=p(V,x)$. This circuit $V'$ applies Cliffords $C_1, C_2$ to both witnesses padded with some number of $\ket0$, then verifies with $V$. In total, $V'$ has proof length at most $4t$.
\end{theorem}

Note that this result is incomparable to that of Theorem \ref{thm:qmak}; even though it is well-known that $\QMA[k]$ can be reduced to $\QMA[2]$ \cite{harrow2013testing}, these reductions increase the number of $T$ gates proportionally to the proof length of the original protocol. %
Whether there exists a magic-preserving reduction or an output-size-independent compilation remains an open question.

For interactive multi-round proofs, we first consider the single-coin $\QMAM$ protocol: Merlin first sends a quantum register, Arthur responds with a uniformly random classical bit, and Merlin responds with a second quantum register. Marriott and Watrous showed that this simple protocol captures all of $\QIP$ \cite{MarriottWatrous2005} under polynomial-time reductions, making this an intriguing class to study. Note that our bound considers the $T$ gate count of a verifier here to be the maximum number of $T$ gates used in the circuit for outcome $0$ and for outcome $1$. However, as with the $\QMA[k]$ to $\QMA[2]$ reduction, reducing a general $\QIP$ protocol to three messages often increases the magic complexity proportional to the original protocol size. We show the following theorem:%

\begin{theorem}[Informal] \label{thm:qmam}
Let $V_0,V_1$ be the verification circuits in single-coin $\mathsf{QMAM}$ that use at most $t$ $T$ gates in total and measure one output qubit. There exists efficiently computable circuits $V'_0,V'_1$ such that $p((V'_0,V'_1),x)=p((V_0,V_1),x)$ for every input $x$. Furthermore, $V'_0,V'_1$ are simply an application of a Clifford $C_1$ to the first message register, and a Clifford $C_2^{i}$ to the second message based on the coin's outcome $i$ followed by applying $V_i$. Merlin sends at most $2(t+1)$ qubits in the first message and at most $t+1$ qubits in the second, for a total of at most $3(t+1)$ qubits.
\end{theorem}

In the pursuit of a result for general $\QIP$, we also show a special case for $t=0$, when Arthur is composed entirely of non-adaptive Clifford gates. For these protocols, Merlin's optimal strategy can always be expressed as a set of non-adaptive Clifford circuits. This allows us to complete the following:

\begin{theorem}[Informal] \label{thm:qipinp}
In a $\QIP$ protocol where Arthur is composed only of non-adaptive Clifford circuits, Merlin can attain optimal acceptance using only non-adaptive Clifford operations.
Any problem $f\in\QIP$ with Arthur satisfying such constraints belongs to $\sP$.
\end{theorem}

Taken together, these results demonstrate how restrictions on Arthur's ability as a verifier limit the communication required from Merlin to solve interactive problems, and we see these as offering evidence for a positive answer to communication compilations for general $\QIP$.

\subsection{Techniques}

\paragraph{Magic Complexity in Verification.}
To analyze the magic complexity of Arthur, we choose to only count the number of $T$-gates his circuit uses. This is because any circuit can be $\epsilon$-approximated using at most $\polylog(1/\epsilon)$ Clifford gates and $T$-gates \cite{dawson2005solovay}. Thus, we can assume without loss of generality that Arthur's circuit is only composed of this limited gate set.

Each $T$-gate can be implemented by the well-known post-selection gadget pictured below.

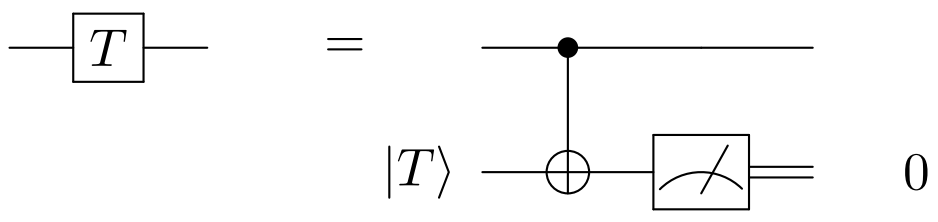
\begin{figure}[H]
    \centering
    \makebox[\textwidth][c]{%
        \ensuremath{\Qcircuit @C=1.2em @R=1em {
            & \gate{T} & \qw
            & \push{\quad=\quad}
            & & \ctrl{1} & \qw & \qw & \\
            & & & & \lstick{\ket{T}}
            & \targ & \meter & \cw & \rstick{0}
        }}%
    }
    \label{fig:magic_injection}
\end{figure}

where $\ket{T}$ is the state $\frac{1}{\sqrt{2}}\ket{0}+\frac{e^{i\pi/4}}{\sqrt{2}}\ket{1}$. The measurement is in the standard basis and post-selected on outcome ${0}$. This occurs with probability $1/2$ regardless of the input $\ket{\psi}$.

If Arthur's original circuit $V$ used $t$ $T$-gates, then replacing each $T$-gate with a post-selection gadget results in a circuit $V_C$ that: uses only Clifford gates, consumes $t$ $\ket{T}$ states, and succeeds in measuring each as $\ket{0}$ with total probability $1/2^t$. Conditioned on these measurements succeeding, $V_C$ produces the same output state as the original, with the measured $\ket{T}$ register appended. This technique allows us to better understand the entanglement between the magic as a communication resource and Merlin's proofs.

If $V$ accepts on measuring one qubit as $\ket{1}$, then $V_C$ accepts by measuring the same qubit as $\ket{1}$ with additional $\ket{0}$ measurements for each of the $t$ $\ket{T}$ states. This accepting observable is a stabilizer projector $\Pi^S$ on $t+1$ qubits of the output state; it acts as $\ketbra{1}{1}$ on the output qubit and $\ketbra{0}{0}^{\otimes t}$ on the $\ket{T}$ states.

$\Pi^{S'}:=V_C^\dagger\Pi^S V_C$ is then the accepting observable acting on the input to Arthur's circuit, which is composed of Merlin's witness, Arthur's ancilla, and the $\ket{T}$ states. Since $\Pi^S$ is a stabilizer projector and $V_C$ is a Clifford circuit, $\Pi^{S'}$ is also a stabilizer projector. This observation gives rise to our two main compilation techniques: extraction and compression. These techniques analyze the behavior of $\Pi^{S'}$ on the witness register, finding gadgets to separate out qubits that can be optimized by Arthur from those that must be communicated by Merlin.

\paragraph{Extraction.}

The technique we coin ``extraction" comes from a wider body of literature concerning itself with the following question: given a stabilizer state $\ket{\psi}$ on $n$ qubits partitioned across $m$ parties, can each party apply a Clifford operation on their qubits to decompose $\ket{\psi}$ into a product of simpler states? Extracting these simpler states out of $\ket{\psi}$ effectively characterizes the entanglement structure of $\ket{\psi}$ across these $m$ party partitions, correlating qubits between parties.

Arunachalam \emph{et. al.}~\cite{ABNO2022} implicitly use this technique in their compilation of $\QMA$: after fixing Arthur's ancilla register to be $\ket{0}$, the resulting observable $\Pi^{S''}$ is a stabilizer projector on two parties: the witness register and the $\ket{T}$ register. Bravyi, Fattel, and Gottesman \cite{BFG06} gave party-local Cliffords to decompose this projector into a tensor product of one- or two-qubit projectors. Furthermore, the one-qubit projectors could be maximized by $\ket{0}$, and the two-qubit projectors correlated one qubit in each party. Since the second party, the $\ket{T}$ register, is at most $t$ qubits (one for each post-selection gadget), at most $t$ qubits of the witness register are correlated with them. The remaining witness qubits can maximize the projector simply by being set to $\ket{0}$. Thus, Merlin only needs to communicate those $t$ qubits; the remainder can be supplied by Arthur.

We generalize this technique in two ways. First, we show that extraction of an $m$-party stabilizer projector is essentially equivalent to extraction of an $m+1$-party stabilizer state. This is done by introducing a purifying register to the projector. Second, we use an observation from Englbrecht, Kraft, and Kraus \cite{EKK2022}. They note that every indecomposable state produced by a 4-party extraction must be `balanced' across the parties; if one party holds $n$ qubits of an indecomposable state, no other party can hold more than $2n$ qubits of that state. This can be recursively applied to 3-party projectors, yielding the extraction technique: for any stabilizer projector partitioned across three parties, there exist efficiently computable party-local Clifford operations decomposing it into a tensor product of `balanced' projectors.

This technique can be immediately applied to $\QMA[2]$. After fixing the ancilla register, the observable $\Pi^{S''}$ is a stabilizer projector on three parties: the two unentangled witness registers, and the $\ket{T}$ register. Extracting $\Pi^{S''}$ reveals that at most $2t$ qubits of each witness can be correlated with the $\ket{T}$ register. A careful analysis of the terms acting on the remaining qubits confirms that they can be maximized by supplying $\ket{0}$, which Arthur can provide. Thus, after compiling Arthur to apply these extracting gadgets, Merlin only needs to communicate at most $4t$ qubits.

However, this technique does not generalize for $\QMA[3]$ or beyond. With $k$ unentangled proofs, the witnesses register must be partitioned into $k$ parties to maintain the unentanglement of witnesses. However, the `balanced' observation from \cite{EKK2022} is not known to hold for four-or-more party stabilizer projectors; indecomposable 4-party projectors could seemingly correlate an arbitrary number of witness qubits to very few $\ket{T}$ qubits. For an arbitrary number of parties, we develop the following technique.

\paragraph{Compression.}

For $\QMA[k]$, the technique of compression simply treats each witness, the ancilla, and the $\ket{T}$ register as separate registers. It then bounds the nontrivial support of $\Pi^{S'}$ on the $k$ witness registers through register-local Clifford operations. We do this by analyzing the rank of the stabilizer group of $\Pi^{S'}$.

First, we show that, for any set of $m$ Pauli strings acting on $n$ qubits, there is an efficiently computable Clifford change of basis that sends it to a set of $m$ Pauli strings whose action is supported on at most $m$ qubits. Notably, this means that, under this change of basis, $n-m$ qubits are acted on trivially by all $m$ Pauli strings. As an example, a single Pauli string like $Z_1\otimes Z_2\otimes ...\otimes Z_n$ acts on $n$ qubits, but applying a Clifford sends this to a parity check on a single qubit and identity on the remainder: $Z_1\otimes \II_2 \otimes...\otimes \II_n$. Effectively, this Clifford `compresses' the action of the set onto few qubits.

This can then be applied to the stabilizer group of $\Pi^{S'}$, $S'$. Since $\Pi^{S}$ is simply $t+1$ independent Pauli checks, its stabilizer group $S$ has rank $t+1$. This rank is preserved by Clifford conjugation, so $S'$ also has rank $t+1$. This means that it is generated by a set of $t+1$ Pauli strings. We then apply compression to each register independently. This produces a register-local Clifford change of basis for each of the $k$ witnesses. Under this basis, in each register, the portion of the generating Pauli strings contained in that register is compressed to at most $t+1$ qubits. Thus, the generating set as a whole acts nontrivially on at most $k(t+1)$ qubits of the witness registers. Since these Pauli strings generate the stabilizer group of the accepting observable, under this change of basis, $\Pi^{S'}$ acts trivially in the witness registers on all but those $k(t+1)$ qubits. Merlin only needs to communicate the optimal witness on these qubits; the remaining witness qubits can be set to any state and still maximize the acceptance observable. For simplicity, we set these to be $\ket{0}$, which Arthur can provide.

Compression can also be applied to $\QMAM$, the restricted form of $\QIP$. Since Arthur's message consists of a classical random bit, we can treat the entire $\QMAM$ protocol as two independent computations: $V'_0$, where the random bit is $0$; and $V'_1$, where the random bit is $1$. In this perspective, Merlin prepares two different witnesses $\rho_0,\rho_1$, one for each outcome. These witnesses are partitioned into a first-message register and a second-message register, and they must agree on the first-message register. If $V'_0,V'_1$ use at most $t$ $\ket{T}$ states in total and measure one output qubit, then their corresponding $\Pi^{S'}_0$, $\Pi^{S'}_1$ will be generated by at most $t+1$ qubits each. The second-message register of each can be independently compressed, bounding it by $t+1$ qubits. The first-message register must satisfy the checks from both computations. This can be done by taking the union of the two generating sets, yielding at most $2(t+1)$ generating Pauli strings. Compressing the first message limits its length to $2(t+1)$ qubits. Thus, the total communication from Merlin is upper-bounded by $3(t+1)$ qubits.

Crucially, since the rank of $S'$ scales with the number of Pauli checks on the output state, compression relies on the post-selection gadgets to contribute the majority of the generating stabilizer set of $V_C$'s acceptance observable. If Arthur measures $o$ qubits instead of 1, then $\Pi^{S}$ will have a stabilizer group with rank $t+o$ instead. Compression then produces a communication bound of $k(t+o)$ qubits for $\QMA[k]$ and $3(t+o)$ qubits for $\QMAM$.

Even though this technique captures $\QMAM$, it fails for general $\QIP$. This is because compression does not characterize Arthur's messages, which could require many-qubit responses even with low magic complexity. However, when this circuit is purely non-adaptive Clifford gates, the following technique perfectly captures the interactive protocol.

\paragraph{Non-Adaptive Clifford Choi Representations for QIP.}

Our results for $\QMA[k]$ tell us that to understand the parameterized complexity of quantum verification by magic, we need to better understand stabilizer projectors. For quantum interactive proofs, $\QIP$, the story is more nuanced, but similar. We build off the techniques of \cite{Gutoski2006}, which introduce the Choi representations of quantum games between two parties, to understand these interactive protocols as a multiplication of two matrices, rather than the awkward alternative of each party's action in the original definition. In this light, we can prove that representations of fully Clifford Arthur and Merlin are proportional to stabilizer projectors $\Pi^A,\Pi^M$.

These representations give us the surprising result that for optimal success of a Clifford Arthur, Merlin only needs to use non-adaptive Cliffords. By representing our optimization as a dual linear program over Merlin's matrix, we are able to decompose the group $A$ of Arthur's projector $\Pi^A$ into degrees of freedom Merlin can contribute to through Gaussian elimination. This makes it easy to find a good candidate for the optimal probability and show a Clifford Merlin that achieves this.

By proving this, we reduce the problem of Arthur finding an optimal transcript for Merlin to a polynomial-sized linear program over the symplectic linear algebra for stabilizer groups. With completeness of $2/3$, we know that for fully Clifford circuits this is equivalent to perfect completeness, and so we must determine if Merlin can provide Cliffords such that the output is stabilized by $Z_1$.

Assuming Arthur measures one output qubit, commutation is preserved for us, and so we simply choose any Pauli string for Merlin at each round that is still invertible---i.e. it does not send $\II$ to a non-identity string or vice versa. However, we further make the observation that there always exists a Merlin that keeps his register non-identity as long as Arthur will allow. This task can then be expressed as a linear program of polynomially many equations and unknowns over $\mathbb{F}_2$, which completes the result.

The major barrier to proving communication compilations for general $\QIP$ seems to be that Arthur's responses introduce a larger number of Pauli checks than for the other verification tasks. We suspect that if such compilations exist, they must involve more complicated gadgets, such as gadgets that correlate across rounds to allow Arthur more memory.

\section{Preliminaries}
In this section we introduce the mathematical formalism behind magic complexity, symplectic linear algebra, Clifford+T compilations, Choi representations, magic state injection, our target complexity classes, and representations for $\QIP$.

\subsection{Stabilizers}
\begin{definition}[Pauli Group]
    The Pauli group $\cP_1$ on one qubit is the subgroup of the unitary group $\cU(2)$ generated by the operators $\cP_1=\langle X,Y,Z\rangle$,
    with
    \begin{equation}
        X^2=Y^2=Z^2=1,\qquad XZ=-ZX,\qquad Y=iXZ.
    \end{equation}
    The $ n$-qubit Pauli group $\cP_n$ is the subgroup of the unitary group $\cU(2^n)$ consisting of tensor products of single-qubit Paulis, or equivalently,
    \begin{equation}
        \cP_n=\{cP_1\otimes\dots\otimes P_n\mid c\in\{\pm1,\pm i\},P_i\in\{X,Y,Z\}\}.
    \end{equation}
    We denote a single-qubit Pauli $P$ acting on the $j$th qubit as $P_j$.
\end{definition}

As a consequence, every Pauli $P$ has eigenvalues either $\pm1$ or $\pm i$ and every pair of Paulis $P,Q$ either commute or anticommute.

\begin{definition}[Clifford Group]
    The Clifford group $\cC_n\leq\cU(2^n)$ is the normalizer of the Pauli group $\cP_n$, meaning
    \begin{equation}
        \cC_n=\{C\in\cU(2^n)\mid C\cP_nC^\dagger=\cP_n\}.
    \end{equation}
    This group is generated by the operators $\cC_n=\langle\CNOT,H,S\rangle$.
\end{definition}

\begin{definition}[Stabilizers]
    A stabilizer group $S$ is an abelian subgroup of the Pauli group $\cP_n$ not containing $-\II$. The rank of this stabilizer group $\rank(S)$ is defined to be the minimum number of elements needed to generate this group.
\end{definition}

Notice that by ensuring that this subgroup is abelian and doesn't contain $-\II$, there must be a non-trivial simultaneous $+1$ eigenspace of the elements of $S$.

\begin{definition}[Stabilizer Projectors]\label{stabproj}
    A stabilizer projector $\Pi^S$ is a projector that can be written in the form
    \begin{equation}
        \Pi^S=C^\dagger(\ketbra{0^{k}}{0^{k}}\otimes \II_{2^{n-k}})C
    \end{equation}
    for some Clifford $C$, and $k=\rank S$. Equivalently, a stabilizer projector is a projector that projects onto the $+1$ eigenspace of some stabilizer group $S$. Indeed, taking $S$ to be the group obtained by conjugating $\langle Z_1,\dots,Z_k\rangle$ by the Clifford $C$ shows that they are indeed equivalent.
\end{definition}

\begin{fact}\label{stabprojrank}
    Let $S\leq\cP_n$ be a stabilizer group and let $\Pi^S$ be its corresponding stabilizer projector. Then $\rank(\Pi^S)=\Tr(\Pi^S) = 2^{n-\rank S}$.
\end{fact}

\begin{fact}\label{stabprojform}
    A stabilizer projector $\Pi^S$ with stabilizer group $S=\langle P_1,...,P_k \rangle$, $k=\rank S$, can be written as
    \begin{equation}
        \Pi^S=\prod_{P \in \{P_1,..,P_k\}} \frac{\II+P}{2} = \frac{1}{2^k}\sum_{P \in S} P.
    \end{equation}
\end{fact}
\begin{fact}\label{traceoutstabproj}
    For any stabilizer projector $\Pi^S$ and register $A$,
    \begin{equation}\Tr_A(\Pi^S) = 2^{|A| + \rank T - \rank S}\Pi^T\end{equation}
    for some stabilizer projector $\Pi^T$.
\end{fact}
\begin{definition}[Purifications of Stabilizer Projectors]\label{purestabproj}
For any stabilizer projector
\begin{equation}
    \Pi^S=C^\dagger(\ketbra{0^k}{0^k}\otimes\II_{2^{n-k}})C
\end{equation}
on register $A$, we say the quantum state $\ket{\psi}_{AB}$ acting on registers $A,B$ is a purification over $B$ if
\begin{equation}
    \Tr_B(\ketbra{\psi}{\psi}) = 2^{|B| + \rank S - n} \Pi^S
\end{equation}
Furthermore,
\begin{equation}
    \ket{\psi} := (C \otimes \II_B)(\ket{0}^{\otimes \rank S} \otimes \ket{\EPR}^{\otimes (n - \rank S)}),
\end{equation}
where $\ket{\EPR}^{\otimes (n - \rank S)}$ are EPR pairs each with one qubit in $A$ and one in $B$, is a valid purification of the projector with purifying register $B$, and is a stabilizer state.
\end{definition}

\begin{fact}\label{stabstateonstabproj}
    For any stabilizer projector $\Pi^S$ and stabilizer state $\ket{\psi}$,
    \begin{equation}\Pi^S\ket{\psi} = 2^{-\alpha/2}\ket{\phi}\quad\text{or}\quad \Pi^S\ket{\psi} = 0\end{equation}
    for some stabilizer state $\ket{\phi}$ and $\alpha \in \mathbb{Z}_{\ge 0}$.
\end{fact}

\subsection{Symplectic vector spaces}
Let $V$ be a vector space over some field $k$. A bilinear form is a map $\omega:V\times V\to k$ that is linear in both arguments. The most common example of a bilinear form is the inner product. Inner products require the form to be conjugate-symmetric, so $\omega(v,w)=\overline{\omega(w,v)}$,
and positive-definite, so $\omega(v,v)>0$
when $v\neq0$. However, one can instead consider alternating forms, which require that $\omega(v,v)=0$
for all $v$. This in turn implies that
\begin{equation}
    \omega(v,w)=-\omega(w,v).
\end{equation}
A bilinear form $\omega$ is said to be non-degenerate if having $\omega(v,w)=0$ for all $w\in V$ implies that $v=0$. If in addition $\omega$ is alternating, then we say the form is symplectic, and the underlying vector space $V$ is a symplectic vector space.
\\\\
Symplectic vector spaces have an important relation with the Pauli group. Recall that, after modding out the center $Z(\cP_n)$ to remove global phase, the Paulis $P\in\cP_n/Z(\cP_n)$ can be written as $P\sim X^xZ^z$ with $x,z\in\bits^n$ and
\begin{equation}
    X^x=\prod_{j\in[n]}X_j^{x_j},\qquad Z^z=\prod_{j\in[n]}Z_j^{z_j}.
\end{equation}
In addition, given two paulis $P\sim X^{x_1}Z^{z_1},Q\sim X^{x_2}Z^{z_2}$,
\begin{equation}
    PQ\sim X^{x_1}Z^{z_1}\cdot X^{x_2}Z^{z_2}\sim X^{x_1+x_2}Z^{z_1+z_2}.
\end{equation}
This implies that there is an isomorphism between vector spaces $\cP_n/Z(\cP_n)\cong\ZZ_2^{2n}$. However, this isomorphism loses information about the commutation relations between the Paulis, which is important data that we would like to keep. So, we introduce a form $\omega:\cP_n\times\cP_n\to\ZZ_2$, with
\begin{equation}
    PQ=(-1)^{\omega(P,Q)}QP.
\end{equation}
One can easily check that this form is bilinear, and since $P$ commutes with itself, $\omega$ is alternating as well. In addition, after modding out by the center of $\cP_n$, $\omega$ is in fact symplectic. This implies that subgroups $S$ of the Pauli group correspond to subspaces $V^S\leq\ZZ_2^n$, and since Clifford unitaries preserve the commutation relation of Paulis, as
\begin{equation}
    CPC^\dagger\cdot CQ C^\dagger=CPQC^\dagger=(-1)^{\omega(P,Q)}CQPC^\dagger=(-1)^{\omega(P,Q)}CQC^\dagger\cdot CPC^\dagger,
\end{equation}
the Cliffords $C$ correspond to isomorphisms $T$ preserving the symplectic form $\omega$.

\subsection{Magic state injection}
Define the $T$ gate to be the operator in $\cL(\CC^2)$ given by the following matrix,
\begin{equation}
    T=\begin{pmatrix}
        1&0\\0&e^{i\pi/4}
    \end{pmatrix}.
\end{equation}
One can see that the $T$ gate is not a Clifford unitary, but Cliffords and $T$ gates can $\epsilon$-approximate any two-qubit unitary in $\polylog(1/\epsilon)$ many gates \cite{dawson2005solovay}. In addition, the Gottesman-Knill theorem \cite{Gottesman1999} states that an $n$-qubit quantum circuit $U$ consisting of $m$ Clifford gates and $t$ $T$ gates can be simulated classically in $\poly(n,m)2^{O(t)}$ time. So it is therefore useful to parametrize circuits by their ``magic'' complexity, the number of $T$ gates needed to compute the circuit.
\\\\
Define the magic $T$ state $\ket{T}\in\CC^2$ to be the state
\begin{equation}
    \ket T=T\ket+=\frac{\ket0+e^{i\pi/4}\ket1}{\sqrt{2}}.
\end{equation}

\begin{fact}[Magic State Injection]
    In order to apply the unitary $T$ to a state $\ket{\psi}$, we can first supply the state $\ket{T}$ on a separate qubit and perform
    \begin{equation}
        \CNOT_{R \rightarrow M} \ket{\psi}_R\ket{T}_M,
    \end{equation}
    and then apply a standard basis measurement on the $M$ register. On outcome 0, which occurs with probability $1/2$, the resulting state will be $T\ket{\psi}\ket{0}$.
\end{fact}

This implies that we can view the circuit $U\ket\psi$ with $U$ containing $t$ $T$ gates as $C\ket\psi\ket*{T^t}$ post-selected on measuring the last register and receiving $0^t$, with Clifford $C$.

\subsection{Complexity Classes Definitions}

In this section, we formally review the complexity classes we work with throughout the paper and the Clifford+T compilation used to understand their magic complexity.

It is known from the work of \cite{Boykin2000} on quantum gate universality and the Solovay-Kitaev Theorem \cite{Kitaev1997} that any quantum circuit has a universal gate set $\{T,\CNOT,H\}$ and that any polynomial-sized quantum circuit can be approximated to inverse polynomial accuracy with at most polynomial overhead in size using this gate set. These two facts give a nice recompilation of any of our standard complexity classes as Clifford+T circuits with standard constants.

\begin{definition}[Non-Adaptive Clifford+T Circuits]
    A non-adaptive Clifford+T circuit is one in which the only gates available are Cliffords and the T gate defined above. Non-adaptive means that no mid-circuit measurements or classical conditioning are allowed.
\end{definition}

For the remainder of the paper, we will assume all circuits are non-adaptive Clifford+T circuits. Using this, we define a parameterized version of our quantum verification classes as follows.
\begin{definition}[$\QMA_t{[k]} (c,s,o)$]
    A function $f:\{0,1\}^*\rightarrow\{0,1\}$ belongs to
    $\QMA_t[k](c,s,o)$ if there exists a polynomial $p$ and a
    polynomial-time uniform family of quantum circuits $\{V_n\}_{n\in\NN}$, where $V_n$ uses at most $t$ T gates, satisfying the following conditions.
    \medskip
    
    \noindent\textbf{The Protocol.}

    On input $x\in\{0,1\}^n$, Arthur receives messages $\ket{m_1},...,\ket{m_k}$ from Merlin and the circuit in Figure \ref{fig:qmak_circuit} is run.
\begin{figure}[H]
    \centering
    \makebox[\textwidth][c]{%
        \Qcircuit @C=1.5em @R=1.5em {
        \lstick{\ket{m_1}}
            & \qw & \qw & \multigate{4}{V_{n}}
            & \qw & \meter & \\
        \lstick{\vdots}
            & & & \nghost{V_{|x|}}
            & \push{\vdots} &
            & \\
        \lstick{\ket{m_k}}
            & \qw & \qw & \ghost{V_{n}}
            & \qw & \qw & \\
        \lstick{\ket{x}}
            & \qw & \qw & \ghost{V_{n}}
            & \qw & \qw & \\
        \lstick{\ket{0^{p(n)-n}}}
            & \qw & \qw & \ghost{V_{n}}
            & \qw & \qw &
        }
    }%
    \caption{The $\mathrm{QMA}[k]$ Protocol.}
    \label{fig:qmak_circuit}
\end{figure}
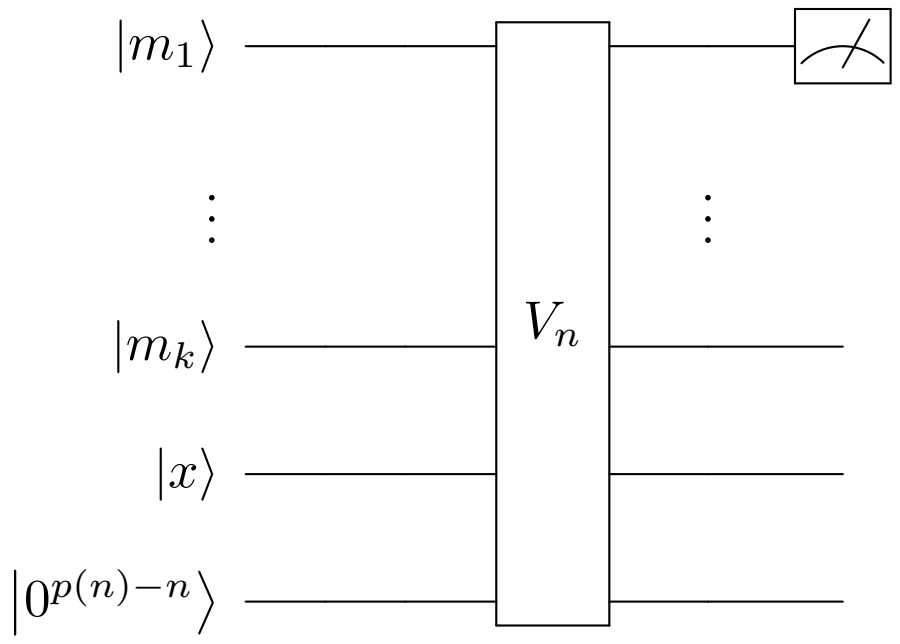%
    Let $p_{A,x}(\ket{m_1},\ldots,\ket{m_k})$
    denote the probability that the circuit measures $0$ on all outputs on input $x$
    against the Merlin messages. The following conditions must hold for every
    $x\in\{0,1\}^*$.
    \begin{itemize}
        \item \textbf{Completeness.}
        If $f(x)=1$, then there exists a Merlin strategy
        $(\ket{m_1},...,\ket{m_k})$ such that
        \begin{equation}
            p_{A,x}(\ket{m_1},...,\ket{m_k})
            \ge c.
        \end{equation}

        \item \textbf{Soundness.}
        If $f(x)=0$, then for every Merlin strategy
        $(\ket{m_1},...,\ket{m_k})$,
        \begin{equation}
            p_{A,x}(\ket{m_1},...,\ket{m_k})
            \le s.
        \end{equation}
    \end{itemize}
\end{definition}
\begin{definition}[$\QIP_t{[k]}(c,s,o)$]
    A function $f:\{0,1\}^*\rightarrow\{0,1\}$ belongs to
    $\QIP_t[k](c,s,o)$ if there exists a polynomial $p$ and a
    polynomial-time uniform family of circuits $\{(A_1^n,\ldots,A_k^n)\}_{n\in\mathbb N},$
    where each $A_i^n$ is a circuit acting on $2p(n)$
    qubits and the sum of T gates across all $A_i^n$ is at most $t$, satisfying the following conditions.

    \medskip
    \paragraph{The Protocol.} For an input $x\in\{0,1\}^n$, Merlin chooses $k$ unitaries $\Phi_1,...,\Phi_k$ and the circuit in Figure \ref{fig:qip_circuit} is run.

\begin{figure}[H]
    \centering
    \makebox[\textwidth][c]{%
        \Qcircuit @C=1.2em @R=1.5em {
        \lstick{\ket{0^{2kp(|x|)}}}
            & \multigate{1}{\Phi_1} & \qw
            & \multigate{1}{\Phi_2} & \qw
            & \qw & \push{\hspace{0.5em}\cdots\hspace{0.5em}} &
            & \multigate{1}{\Phi_k}
            & \qw & \qw & & \\
        \lstick{\ket{0^{p(|x|)}}}
            & \ghost{\Phi_1} & \multigate{2}{A_1^{n}}
            & \ghost{\Phi_2} & \multigate{2}{A_2^{n}}
            & \qw & \push{\cdots} &
            & \ghost{\Phi_k} & \multigate{2}{A_k^{n}}
            & \meter & & \\
        \lstick{\ket{x}}
            & \qw & \ghost{A_1^{n}}
            & \qw & \ghost{A_2^{n}}
            & \qw & \push{\cdots} &
            & \qw & \ghost{A_k^{n}}
            & \qw & & \\
        \lstick{\ket{0^{p(n)-n}}}
            & \qw & \ghost{A_1^{n}}
            & \qw & \ghost{A_2^{n}}
            & \qw & \push{\cdots} &
            & \qw & \ghost{A_k^{n}}
            & \qw &
            & \\
        }
    }
    \caption{The $\QIP$ protocol.}
    \label{fig:qip_circuit}
\end{figure}
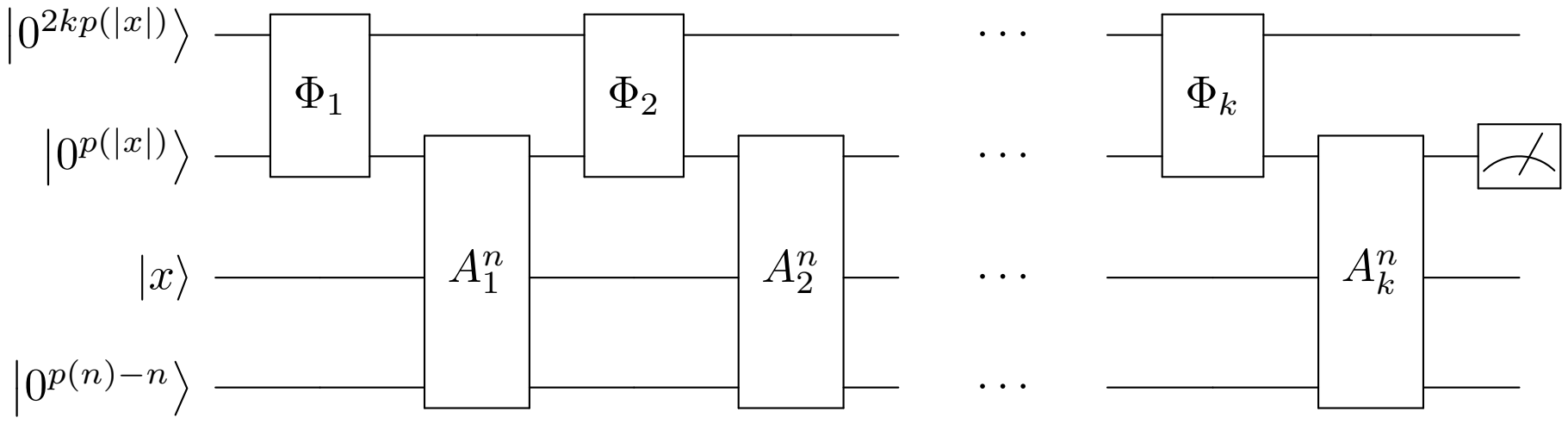%

    \medskip
    \noindent
    Let
    $p_{A,x}(\Phi_1,\ldots,\Phi_{k})$
    denote the probability that the circuit measures $0$ on every output qubit on input $x$
    against Merlin's unitaries. The following conditions must hold for every
    $x\in\{0,1\}^*$:
    \begin{itemize}
        \item \textbf{Completeness.}
        If $f(x)=1$, then there exists a Merlin strategy
        $(\Phi_1,\ldots,\Phi_{k})$ such that
        \begin{equation}
            p_{A,x}(\Phi_1,\ldots,\Phi_{k})
            \ge c.
        \end{equation}

        \item \textbf{Soundness.}
        If $f(x)=0$, then for every Merlin strategy
        $(\Phi_1,\ldots,\Phi_{k})$,
        \begin{equation}
            p_{A,x}(\Phi_1,\ldots,\Phi_{k})
            \le s.
        \end{equation}
    \end{itemize}
\end{definition}

\begin{definition}($\QMAM_t(c,s)$ \cite{MarriottWatrous2005})\label{def:qmam}
    We define $\QMAM_t(c,s)$ similarly to $\QIP_t[k](c,s,o)$ but with the protocol having $k=2$ rounds and after Merlin's first message, Arthur prepares an EPR pair, called the coin, and sends one half to Merlin. After receiving Merlin's second message, he measures the coin and conditionally applies a circuit $V_0$ or $V_1$ with at most $t$ T gates total between them and measures his output qubit.
\end{definition}

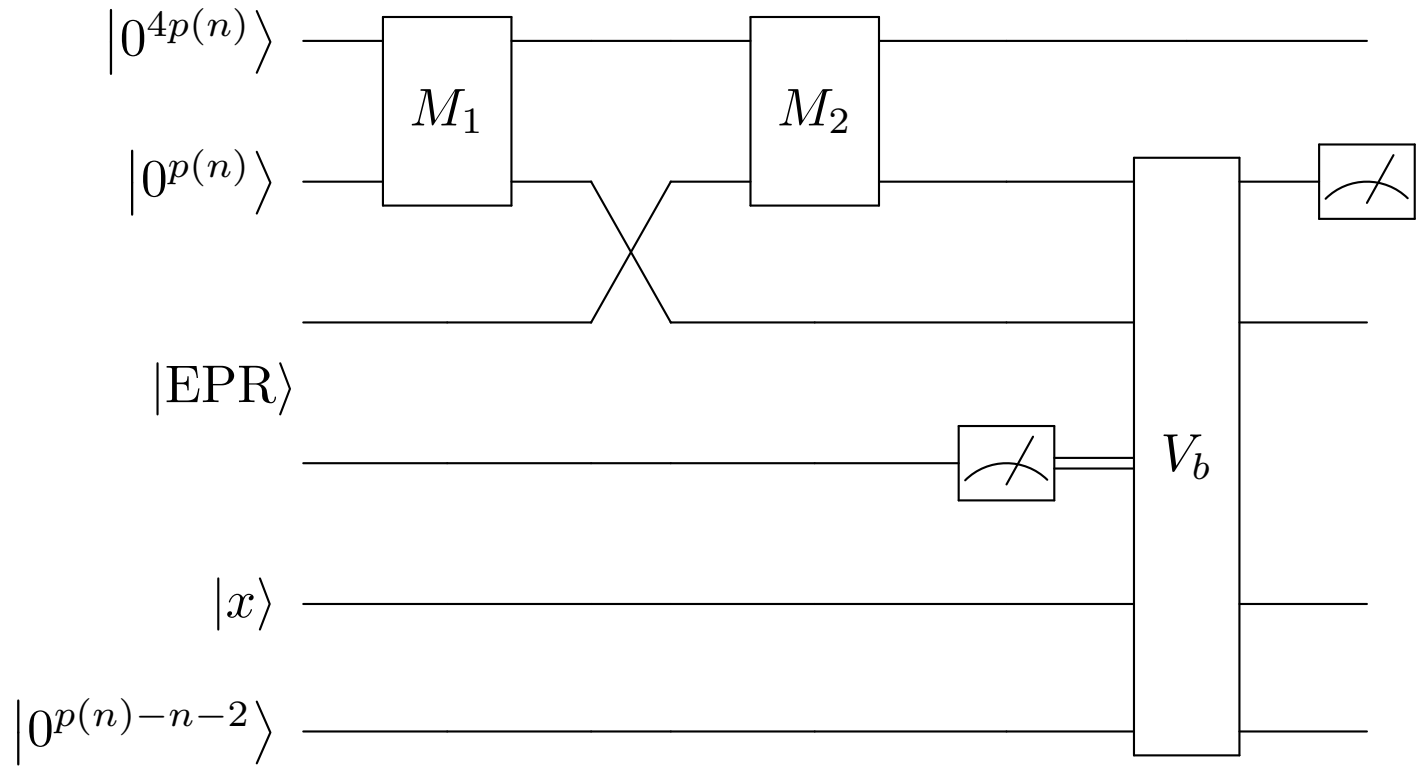
\begin{figure}[H]
    \centering
    \makebox[\textwidth][c]{%
        \Qcircuit @C=1.5em @R=1.5em {
        \lstick{\ket{0^{4p(n)}}}
            & \multigate{1}{M_1} & \qw & \qw
            & \multigate{1}{M_2} & \qw & \qw & \qw \\
        \lstick{\ket{0^{p(n)}}}
            & \ghost{M_1} & \qw \link{1}{1} &
            & \ghost{M_2} & \qw
            & \multigate{4}{V_b} & \meter \\
        \lstick{}
            & \qw & \qw \link{-1}{1} &
            & \qw & \qw
            & \ghost{V_b} & \qw \\
         \lstick{}
            & \qw & \qw & \qw
            & \qw & \meter
            & \cghost{V_b}
              \inputgroup{3}{4}{1.25em}{\ket{\EPR}\quad} \\
        \lstick{\ket x}
            & \qw & \qw & \qw
            & \qw & \qw
            & \ghost{V_b} & \qw \\
        \lstick{\ket{0^{p(n)-n-2}}}
            & \qw & \qw & \qw
            & \qw & \qw
            & \ghost{V_b} & \qw
        }
    }
    \caption{The QMAM Protocol.}
    \label{fig:qmam}
\end{figure}

\subsection{Choi Representations of QIP}

The task of optimizing over Merlin's actions in interactive protocols can be awkward in the definition above, as operationally we have an alternation between applying Arthur's matrices, followed by Merlin. We introduce Choi representations as a solution to this problem in our paper.

Choi representations originally were used to answer a natural question in quantum information: is it possible to somehow embed the information about a unitary transformation into a quantum state that we can then send to someone? They are precisely a way to do so.

\begin{definition}[Choi Representation of a Unitary]\label{choi}
For any unitary matrix $A \in \mathcal{U}(2^n)$, the unnormalized Choi representation, denoted $J(A)\in\CC^{2^n}\otimes\CC^{2^n}$, is defined as
\begin{equation}J(A) = 2^{n/2}(A \otimes \II)\ket{\Omega},\end{equation}
where $\ket{\Omega}$ is $n$ tensored copies of the EPR state, the $i$-th one being between the $i$-th qubit of the left register and the $i$-th qubit of the right register.
\end{definition}

Structurally, the Choi representation is simply a way to embed the matrix entries of $A$ into a column vector. Intuitively, the state is the maximally entangled state between two parties that would be used in the canonical quantum teleportation protocol. We now explore how we can apply Choi representations to quantum interactive proof systems using the quantum games framework introduced in \cite{Gutoski2006}. Our choice of QIP definition above makes it easy to view the protocol as a quantum game. In this view, Arthur is simply a measuring strategy, and Merlin is a non-measuring co-strategy. We then apply their Choi representation methods to embed the entirety of their strategies into single matrices.

\begin{definition}[Measuring Strategies and Non-Measuring Co-Strategies]\label{stratcostrat}
    Define
    \begin{equation}\cS_0(M_{0},A_{0}) := \coS_0(M_{0},A_{0}) := \{1\}.\end{equation}
    More generally, let $\cS_k(M_{1..k},A_{1...k})$ be the set of all non-measuring strategies defined in \cite{Gutoski2006}, that is, the set of all positive matrices $P$ acting on the register $M_{1...k} \otimes A_{1...k}$ that satisfy the following condition
    \begin{equation}\Tr_{A_{k}}(P) = P' \otimes \II_{M_k},\end{equation}
    where $P' \in \cS_{k-1}(M_{1...k-1},A_{1...k-1})$.
    \\\\
    Let $\coS_k(M_{1...k},A_{1...k})$ be the set of all non-measuring co-strategies defined in \cite{Gutoski2006}, that is, the set of all positive matrices, $P$, acting on the register $M_{1...k} \otimes A_{1...k}$ that satisfy the following condition
    \begin{equation}P = P' \otimes \II_{A_{k}}\end{equation}
    where $\Tr_{M_k}(P') \in \coS_{k-1}(M_{1...k-1},A_{1...k-1})$.
\end{definition}

\begin{definition}[Measuring Strategies]
    Define $\Acc_k(M_{1...k},A_{1...k})$ to be the set of all two-outcome measuring strategies in \cite{Gutoski2006}, that is, the set of all tuples of positive matrices, $(P_0,P_1)$, acting on the register $M_{1...k} \otimes A_{1...k}$ such that there exists $P \in \cS_k(M_{1...k},A_{1...k})$ satisfying
    \begin{equation}P = P_0 + P_1 \end{equation}
\end{definition}

These classes are useful because they exactly represent possible strategies of Arthur and Merlin. We now show how to derive the Choi representations of both parties.

\begin{definition}[Representations of Arthur and Merlin]\label{representation}
 For any Arthur and Merlin defined in $\QIP$, we define two maps, which we will call the Choi representations of Arthur and Merlin.

 The construction of these representations is done by finding the map that takes all the opposite party's messages together and outputs the current party's messages and memory. We then find the Choi representation of this and trace out the memory.
    \\\\
    For Merlin, we'll call this mapping $B:A_{1...k-1} \rightarrow M_{1...k} \otimes \Memory$ where $M$ is Merlin's messages, $A$ is Arthur's messages, and $\Memory$ is Merlin's memory.
    \begin{equation}B \coloneqq (\Phi_{k} \otimes \II_{M_{1\dots k-1}})\dots(\II_{A_{2\dots k-1}} \otimes \Phi_2 \otimes \II_{M_1})\qty(\II_{A_{1\dots k-1}} \otimes \Phi_1 \ket{0^{(2k+1)p(|x|)}}_{M_1\Memory})\end{equation}
    Now we apply our Choi representation to this to get
    \begin{equation}J_M(\Phi_{1},\dots,\Phi_{k}) \coloneqq \Tr_{\Memory}(J(B)J(B)^\dagger)^\top \otimes \II_{A_k},\end{equation}
    where we will call $J_M(\Phi_{1},\dots,\Phi_{k})$ the representation of Merlin. Note that the EPR pairs are paired between $A_{i},M_{i+1}$ for $1 \le i \le k-1$.
    \\\\
    We apply this same process to generate a map for Arthur, but additionally we apply our projective measurement $\{P_{\acc}:= \ketbra{0^o}{0^o} \otimes \II,\II-P_{\acc}\}$. We'll define the mapping $W:M_{1\dots k} \rightarrow A_{1\dots k} \otimes \Memory$ where $\Memory$ is Arthur's memory.
    \begin{equation}W \coloneqq \sqrt{P_{acc}}(A_{k} \otimes \II_{A_{1\dots k-1}})\dots(\II_{M_{2\dots k}} \otimes A_1)\qty(\II_{M_{1\dots k}} \otimes \ket{x}\ket{0^{p(|x|-|x|)}}_{\Memory})\end{equation}
    \begin{equation}J_A(P_{\acc},A_1,\dots,A_k) \coloneqq \Tr_{\Memory}(J(W)J(W)^\dagger)\end{equation}
    where we will call $J_A(P_{\acc},A_1,\dots,A_k)$ the representation of Arthur. Note that the EPR pairs are paired between $M_{i},A_i$ for $1 \le i \le k$. For technical reasons later on in the paper, we additionally defined a non-measuring Arthur, denoted $J_{\tilde A}(A_1,\dots,A_k)$ as the same map without the projection in $W$.
\end{definition}
\begin{fact}[Complete Characterizations of Representations \cite{Gutoski2006}]\label{repiff}
    We can now use these constructions to provide a classification of these representations defined above.
    \begin{enumerate}
        \item A matrix $A_{\acc}$ is contained in an element of $\Acc_k(M_{1...k},A_{1...k})$ if and only if there exists an Arthur, described by $P_{\acc},A_1,...,A_k$, such that $J_A(P_{\acc},A_1,...,A_k) = A_{\acc}$.
        \item A matrix $A \in \cS_k(M_{1...k},A_{1...k})$ if and only if there exists a non-measuring Arthur, described by $A_1,...,A_k$, such that $J_{\tilde A}(A_1,...,A_k) = A$.
        \item A matrix $M\in \coS_k(M_{1...k},A_{1...k})$ if and only if there exists some Merlin, described by $\Phi_{1},...,\Phi_{k}$, such that $J_M(\Phi_{1},...,\Phi_{k}) = M$.
    \end{enumerate}
\end{fact}

\begin{fact}[Acceptance Probability of Representations \cite{Gutoski2006}]\label{qipsuccessprobability}
    For any Arthur and Merlin defined in $\QIP[k](c,s,o)$, described by $P_{acc},A_1,...,A_k$ and $\Phi_{1},...,\Phi_{k}$ respectively,

    \begin{equation}p_{A,x}(\Phi_{1},...,\Phi_{k}) = \Tr(J_A(P_{\acc},A_1,...,A_k) \cdot J_M(\Phi_{1},...,\Phi_{k})).\end{equation}
\end{fact}
\section{Communication upper bounds for $\QMA[k]$}

\subsection{$\QMA[2]$: Extraction}
This section applies the extraction technique to show a communication upper bound for $\QMA[2]$. The first goal will be to understand the technique of extraction and how this applies to stabilizer projectors.

\begin{definition}[Extraction]
    Let $\ket\psi$ be a stabilizer state on $n$ qubits, and let $\alpha_1,\dots,\alpha_k$ be some partition of $[n]$ into $k$ parties. We say that a stabilizer state $\ket{\phi}$ can be extracted from $\ket{\psi}$ if there exists Cliffords $C_j$ acting on $\alpha_j$ for $1 \le j \le k$ and some stabilizer state $\ket\gamma$ such that
    \begin{equation}C_1 \otimes ... \otimes C_k \ket{\psi} = \ket{\gamma} \otimes \ket{\phi}\end{equation}
    Similarly, a stabilizer projector $\Pi'$ can be extracted from a stabilizer projector $\Pi$ if there are Cliffords $C_j$ acting on $\alpha_j$ for $1\leq j\leq k$ and some stabilizer projector $\Pi''$ such that
    \begin{equation}(C_1 \otimes ... \otimes C_k)\Pi(C_1^\dagger \otimes ... \otimes C_k^\dagger)=\Pi''\otimes\Pi'.\end{equation}
    We say a stabilizer state or projector is ``decomposable'' if there exists any smaller state that can be extracted from it.
\end{definition}

\begin{lemma}[Projector to State Extraction]\label{lem:projtostateextraction}
    Let $\Pi^\psi$ be a stabilizer projector on $k$ parties $\alpha_1,...,\alpha_k$. Let the stabilizer state $\ket{\psi}$ be a purification with purifying register $A$. Then, a stabilizer projector $\Pi^\phi$ on register $E$ can be extracted from $\Pi^\psi$ if there exists a stabilizer state $\ket{\phi}$ on $E$ extracted from $\ket{\psi}$ on parties $\alpha_1,...,\alpha_k,A$ that is a purification of $\Pi^\phi$ over $A$.
\end{lemma}
\begin{proof}
    If there is some $\ket{\phi}$ that can be extracted, then there exists Cliffords 
    $C_{1},...,C_{k+1}$ acting on $\alpha_1,...,\alpha_k,A$ respectively s.t.
    \begin{equation}
        C_1 \otimes ... \otimes C_k \otimes C_{k+1} \ket{\psi} = \ket{\gamma} \otimes \ket{\phi}
    \end{equation}
    This implies that
    \begin{equation}
    (C_1 \otimes ... \otimes C_k)\Tr_A(\ketbra{\psi}{\psi})(C_1^\dagger \otimes ... \otimes C_k^\dagger)= \Tr_{A_{\gamma}}(\ketbra{\gamma}{\gamma}) \otimes \Tr_{A_\phi}(\ketbra{\phi}{\phi})
    \end{equation}
    where $A = A_\gamma \sqcup A_\phi$. Using the definitions of purifications, applying Fact \ref{traceoutstabproj}, and noting that these are projectors, we get
    \begin{equation}
    (C_1 \otimes ... \otimes C_k)\Pi^\psi(C_1^\dagger \otimes ... \otimes C_k^\dagger) = \Pi^\gamma \otimes \Pi^\phi
    \end{equation}
    for some stabilizer projector $\Pi^\gamma$ and the result holds.
\end{proof}
Next, we make use of Observation 14 from \cite{EKK2022} which proves that sufficiently unbalanced states on four parties are always decomposable. We then relate this to extraction on three-party stabilizer projectors to get our key lemma.
\begin{proposition}[Observation 14 \cite{EKK2022}]\label{prop:observation14}
Let $\ket{\psi}$ be a stabilizer state on four parties $\alpha_1,\alpha_2,\alpha_3,\alpha_4$ with $|\alpha_1| \ge |\alpha_2| \ge |\alpha_3| \ge |\alpha_4|$. If $|\alpha_1| > 2|\alpha_4|$, then the state $\ket{\GHZ_k} := \frac{\ket{0^k} + \ket{1^k}}{\sqrt{2}}$ for some $1 \le k \le 4$ can be extracted from $\ket{\psi}$, where each party holds at most a single qubit of this state.
\end{proposition}

Bridging these two theorems together, we get the following:

\begin{corollary}[Extraction of Unbalanced Projectors] \label{cor:unbalancedprojectorextraction}
Let $\Pi^S$ be a stabilizer projector on three parties $\alpha_1,\alpha_2,\alpha_3$ with $|\alpha_1| \ge |\alpha_2| \ge |\alpha_3|$. If $|\alpha_1| > 2|\alpha_3|$, then at least one of the following can be extracted from it:
\begin{equation}
    \ketbra{0}{0}, \quad \II, \quad \ketbra*{\EPR}, \quad \ketbra{00}{00} + \ketbra{11}{11}, \quad \ketbra*{\GHZ_3},\quad \ketbra{000}{000} + \ketbra{111}{111}
\end{equation}
where each party holds at most a single qubit of the extracted projector.
\end{corollary}
\begin{proof}
    Let $\ket{\psi}$ be the normalized purification of $\Pi^S$ from Definition \ref{purestabproj} with purifying register $A$. Thus $\ket{\psi}$ is a stabilizer state held by parties $\alpha_1,\alpha_2,\alpha_3,A$.
    Notice that
    \begin{equation}|\alpha_L|\geq|\alpha_1|>2|\alpha_3|\geq2|\alpha_S|,\end{equation}
    where $|\alpha_L|$ is the largest party and $|\alpha_S|$ is the smallest party. Then Proposition \ref{prop:observation14} holds and we extract $\ket{\GHZ_k}$. If $k=1$ and the qubit is held by party $A$, we can always keep applying Proposition \ref{prop:observation14} until we get an extraction which has at least one qubit outside of $A$. We can also assume that for $k=1$ the projector is $\ketbra{0}{0}$ by applying a Hadamard $H$ to this qubit. For each case $k=2,3,4$, the remaining projector after tracing out $A$ must be one in the above list, so by Lemma \ref{lem:projtostateextraction}, at least one of these is extractable from $\Pi^S$.
\end{proof}

We now have all of the tools needed to prove a communication compilation for $\QMA[2]$. Using the fact that unbalanced stabilizer projectors decompose with local Cliffords, we can show that only a small portion of Merlin's messages can have complex entanglement with the magic register, thus giving us lots of qubits that are easy for Arthur to optimize himself. The Cliffords used in this extraction can then be used to form a compiled smaller proof verification circuit for Arthur.

\begin{restatethm}{thm:qma2}
    Given some function $f\in \QMA_t[2](2/3,1/3,o)$ with
    verifier $V$, for each input $x$, there are Clifford
    unitaries $C_1,C_2$ acting on the message registers
    $M_1,M_2$, respectively, and $q_i$-qubit states
    $\ket{m_i}$, where $q_i\leq \min\{2t,|M_i|\}$,
    such that an optimal witness for $V$ is given by
    \begin{equation}
        C_1\ket{m_1}\ket*{0^{|M_1|-q_1}}
        \otimes
        C_2\ket{m_2}\ket*{0^{|M_2|-q_2}}.
    \end{equation}
\end{restatethm}

\begin{proof}
    The optimal success probability for the verifier $V$
    on input $x$ is
    \begin{equation}
        \pi(V,x)=\max_{\ket{m_1},\ket{m_2}}
        \|(\ketbra{0^o}{0^o}\otimes\II)
        V\ket{m_1}\ket{m_2}\ket{x}\ket{0^{a}}\|^2.
    \end{equation}
    By applying magic state injection, we can rewrite the
    above success probability as
    \begin{equation}
        \pi(V,x)=2^t\max_{\ket{m_1},\ket{m_2}}
        \|\Pi^Z C\ket{m_1}\ket{m_2}
        \ket{x}\ket{0^a}\ket*{T^t}\|^2,
    \end{equation}
    where $C$ is some Clifford circuit and $\Pi^Z$ is the
    stabilizer projector
    $\ketbra{0^o}{0^o}\otimes\II\otimes\ketbra{0^t}{0^t}$.
    Defining $\Pi^S$ to be the stabilizer projector
    $C^\dagger\Pi^Z C$, we get
    \begin{equation}
        \pi(V,x)=2^t\max_{\ket{m_1},\ket{m_2}}
        \|\Pi^S\ket{m_1}\ket{m_2}
        \ket{x}\ket{0^a}\ket*{T^t}\|^2.
    \end{equation}
    Now using Fact \ref{1,2,1/2}, we can say that
    \begin{equation}
        \pi(V,x)=\gamma 2^{t-r}
        \max_{\ket{m_1},\ket{m_2}}
        \|\Pi^{S'}\ket{m_1}\ket{m_2}\ket*{T^t}\|^2
    \end{equation}
    for some stabilizer projector $\Pi^{S'}$, some
    $0\leq r\leq |x|+a$, and $\gamma\in\{0,1\}$.
    If $\gamma=0$ the result holds trivially, so assume
    $\gamma=1$.

    Let $u$ denote the number of magic qubits remaining
    in the residual projector during extraction. Notice
    that as long as either residual witness register
    contains more than $2u$ qubits, the residual projector
    is decomposable by Corollary
    \ref{cor:unbalancedprojectorextraction}. When $u=0$,
    we can use the bipartite stabilizer projector
    decomposition obtained from the tripartite stabilizer
    state decomposition of \cite{BFG06}.

    Hence repeatedly applying this corollary gives
    Cliffords $C_1,C_2,C_3$ acting on $M_1,M_2$ and the
    $t$ magic states separately such that
    \begin{equation}
        (C_1^\dagger \otimes C_2^\dagger \otimes C_3)
        \Pi^{S'}
        (C_1 \otimes C_2 \otimes C_3^\dagger)
        = \Pi^A \otimes \Pi^M,
    \end{equation}
    where $\Pi^A$ is a product of the extracted projectors
    from Corollary \ref{cor:unbalancedprojectorextraction}
    which act only on the witness registers, and $\Pi^M$
    consists of the residual projector together with all
    extracted projectors which contain magic qubits.

    Notice that if $u$ magic qubits remain in the residual
    projector, then the residual projector acts on at most
    $2u$ qubits from each witness. Furthermore, the extracted
    projectors included in $\Pi^M$ act on at most $t-u$
    additional qubits from each witness, since each such
    projector contains at least one magic qubit and at most
    one qubit from each witness. Thus, $\Pi^M$ contains
    the entire magic register and acts on at most
    \begin{equation}
        2u+(t-u)\leq 2t
    \end{equation}
    qubits from each witness. Let $q_i$ denote the number
    of qubits of witness $i$ contained in $\Pi^M$.
    We can assume these qubits are the first $q_i$ qubits
    of each witness by incorporating local SWAP gates
    into $C_1$ and $C_2$.

    Let $\ket{m_1'}=C_1^\dagger\ket{m_1}$ and
    $\ket{m_2'}=C_2^\dagger\ket{m_2}$. We can now write
    \begin{equation}
        \pi(V,x)=2^{t-r}
        \max_{\ket{m_1'},\ket{m_2'}}
        \|\Pi^A\otimes\Pi^M
        \ket{m_1'}\ket{m_2'}C_3\ket*{T^t}\|^2.
    \end{equation}

    Now we can assume that
    $\ket{m_i'}=\ket{m_i'}_M\otimes\ket{m_i'}_A$,
    where $\ket{m_i'}_M$ is a state on $q_i\leq 2t$
    qubits. Since norms decompose over tensor products,
    we have
    \begin{align}
        &\|\Pi^A\otimes\Pi^M
        \ket{m_1'}\ket{m_2'}C_3\ket*{T^t}\|^2\\
        &=\|\Pi^A\ket{m_1'}_A\ket{m_2'}_A\|^2
        \|\Pi^M\ket{m_1'}_M\ket{m_2'}_M
        C_3\ket*{T^t}\|^2.
    \end{align}

    For any projector in $\Pi^A$ from Corollary
    \ref{cor:unbalancedprojectorextraction}, we claim that
    the maximum over product states is attained at
    $\ket{0}$ for each of its qubits.

    Therefore, Merlin can always achieve his optimal
    success probability by making
    $\ket{m_i'}_A=\ket{0^{|M_i|-q_i}}$, which implies
    there exist $q_i$-qubit states
    $\ket{m_1'}_M,\ket{m_2'}_M$, with $q_i\leq 2t$,
    such that an optimal witness is given by
    \begin{equation}
        C_1\ket{m_1'}_M\ket*{0^{|M_1|-q_1}}
        \otimes
        C_2\ket{m_2'}_M\ket*{0^{|M_2|-q_2}},
    \end{equation}
    completing the proof.
\end{proof}
The strength of this compilation technique for $\QMA[2]$ is that the size of the smaller witness is independent of the number of output qubits. However, the observation from \cite{EKK2022} is only for four-party states and therefore only applies to three-party projectors. It is an open question as to whether unbalanced five-party states and larger can observe similar entanglement structures. Resolving this would allow the same construction to apply to all of $\QMA[k]$.

\subsection{$\QMA[k]$: Compression}

In this section, we will show that any $\QMA[k]$ verification circuit with low magic can be compiled into one also requiring low witness size. Our principal observation in doing so is that the stabilizer projector representing Arthur is high rank. We will start with the following lemma:

\begin{lemma}\label{lem:stab-grp-comp}
    Given a subgroup $S$ of the Pauli group $\cP_n$ generated by $t$ elements $P_1,\dots,P_t\in S$, there is a Clifford unitary $C$ such that $CSC^\dagger$, the subgroup obtained by conjugating each element of $S$ by $C$, has support only on $t$ qubits. Moreover, this Clifford $C$ can be efficiently computed classically in $\poly(n,t)$ time.
\end{lemma}

\begin{proof}
    Recall that this subgroup $S$ can be viewed as a symplectic vector space of dimension at most $t$. There is an efficiently computable canonical basis \cite{CannasDaSilva2001} $\{e_1,\dots,e_k,f_1,\dots,f_k,u_1,\dots,u_\ell\}$ of $S$ with $2k+\ell\leq t$ such that for all $v\in S$ and $i,j$,
    \begin{equation}
        \omega(u_i,v)=0,\qquad\omega(e_i,e_j)=0,\qquad\omega(f_i,f_j)=0,\qquad\omega(e_i,f_j)=\delta_{ij},
    \end{equation}
    where $\delta_{ij}$ is the Kronecker delta, which is 1 when $i=j$ and 0 otherwise. This implies that there is an invertible map $T$ sending $S$ to the space
    \begin{equation}
        S'=\langle Z_1,\dots,Z_k,X_1,\dots,X_k,Z_{k+1},\dots,Z_{k+\ell}\rangle,
    \end{equation}
    which has support on $k+\ell\leq t$ qubits. One can easily see that the commutation relation is preserved, so by Witt's Extension Theorem \cite{EKK2022}, there is a Clifford unitary $C$ sending $S$ to $S'$, as desired.
\end{proof}

\begin{corollary}\label{cor:stab-proj-comp}
    Given a stabilizer projector $\Pi^S$ acting on a $k$-partite space $\cH=\cH_1\otimes\dots\otimes\cH_k$ with $S$ generated by $t$ elements, there are Clifford unitaries $C_1,\dots,C_k$ acting on $\cH_1,\dots,\cH_k$, respectively, such that the stabilizer projector
    \begin{equation}
        \Pi^{S'}=(C_1\otimes\dots\otimes C_k)\Pi^S(C_1^\dagger\otimes\dots\otimes C_k^\dagger)
    \end{equation}
    acts non-trivially on at most $t$ qubits of each party.
\end{corollary}

\begin{proof}
    Consider the space $V^S$ corresponding to subgroup $S\subseteq\cP_n$, and consider the spaces $V^{S_j}=P_jV^S$, where $P_j$ is the projection onto the portion of $V^S$ contained in party $j$. Since for each $j\in[n]$, each $V^{S_j}$ still has dimension at most $t$, apply Lemma \ref{lem:stab-grp-comp} to each $S_j$ to get Cliffords $C_j$ acting on the space $\cH_j$ conjugating each $S_j$ to a group with support only on $t$ qubits. Thus, in total, taking $C=C_1\otimes\dots\otimes C_k$, the stabilizer group $S'=CSC^\dagger$ has support on only $t$ qubits of each party, and hence the stabilizer projector $\Pi^{S'}=C\Pi C^\dagger$ acts nontrivially on $t$ qubits of each party.
\end{proof}

We can now use these results to prove our claim. By following Arunachalam et al. \cite{ABNO2022}, we will show that we can write the success probability of our verification circuit as the norm of a stabilizer projector $\Pi^S$ applied to our input $\ket{m_1}\cdots\ket{m_k}\ket x\ket{0^a}\ket*{T^t}$. In particular, each generator of $S$ comes from measurements and post-selections, so $S$ is only generated by $t+1$ elements. We can therefore apply Clifford unitaries to the witnesses to use Corollary \ref{cor:stab-proj-comp} to give the desired claim.

\begin{restatethm}{thm:qmak}
    Given some function $f\in \QMA_t[k](2/3,1/3,1)$ with verifier $V$, there are Clifford unitaries $C_1,\dots,C_k$ acting on the message registers $M_1,\dots,M_k$, respectively, and at most $(t+1)$-qubit states $\ket{m_1},\dots,\ket{m_k}$ that depend on $x\in\bits^n$ such that the optimal witness for $V$ is given by
    \begin{equation}
        C_1\ket{m_1}\ket*{0^{|M_1|-(t+1)}}\otimes\dots\otimes C_k\ket{m_k}\ket*{0^{|M_k|-(t+1)}}
    \end{equation}
    Furthermore, these Cliffords can be efficiently computed classically in $\poly(n,t)$ time.
\end{restatethm}

\begin{proof}
    The optimal success probability for the verifier $V$ on input $x$ is
    \begin{equation}
        \pi(V,x)=\max_{\ket{m_1},\dots,\ket{m_k}}\|(\ketbra{0}{0}\otimes\II)V\ket{m_1}\cdots\ket{m_k}\ket{x}\ket{0^{a}}\|^2.
    \end{equation}
    By applying magic state injection, we can rewrite the above success probability as
    \begin{equation}
        \pi(V,x)=2^t\max_{\ket{m_1},\dots,\ket{m_k}}\|\Pi^Z C\ket{m_1}\cdots\ket{m_k}\ket x\ket{0^a}\ket*{T^t}\|^2,
    \end{equation}
    where $C$ is some Clifford circuit and $\Pi^Z$ is the stabilizer projector $\ketbra{0}{0}\otimes \II\otimes\ketbra{0^t}{0^t}$. Defining $\Pi^S$ to be the stabilizer projector $C^\dagger\Pi^Z C$, we get
    \begin{equation}\label{eq:acc-prob-stab-anc}
        \pi(V,x)=2^t\max_{\ket{m_1},\dots,\ket{m_k}}\|\Pi^S\ket{m_1}\cdots\ket{m_k}\ket x\ket{0^a}\ket*{T^t}\|^2.
    \end{equation}
    Now notice that $S$ is generated by at most $t+1$ Cliffords. By Corollary \ref{cor:stab-proj-comp}, there exists Clifford unitaries $C_1,\dots,C_k$ such that
    \begin{equation}\Pi^{S'}:=(C_1\otimes\dots\otimes C_k\otimes\II)\Pi^{S}(C_1\otimes\dots\otimes C_k\otimes\II)^\dagger\end{equation}
    acts nontrivially on at most $t+1$ qubits of $\ket{m_j}$ for all $j\in[k]$. Therefore,
    \begin{align}
        \pi(V,x)&=2^t\max_{\ket{m_1},\dots,\ket{m_k}}\|\Pi^{S}(C_1^\dagger\otimes\dots\otimes C_k^\dagger\otimes\II)(C_1\otimes\dots\otimes C_k\otimes\II)\ket{m_1}\cdots\ket{m_k}\ket x\ket{0^a}\ket*{T^t}\|^2\\
        &=2^t\max_{\ket{m_1'},\dots,\ket{m_k'}}\|\Pi^{S'}\ket{m_1'}\cdots\ket{m_k'}\ket x\ket{0^a}\ket*{T^t}\|^2,
    \end{align}
    where $\ket{m_j}=C_j\ket{m_j'}$ for $j\in[k]$. Thus we can assume that $\ket{m_j'}=\ket{m_j''}\ket{0^{|M_j|-t-1}}$, where the $\ket{0}$'s are on the qubits that $\Pi^{S'}$ acts trivially on, and so there is an optimal witness of the form
    \begin{equation}
        C_1\ket{m_1''}\ket*{0^{|M_1|-t-1}}\otimes\dots\otimes C_k\ket{m_k''}\ket*{0^{|M_k|-t-1}},
    \end{equation}
    where $\ket{m_i''}$ are at most $(t+1)$-qubit states. Notice that these Cliffords are indeed efficiently computable, as they are determined by Lemma \ref{lem:stab-grp-comp} which is efficiently computable.
\end{proof}
\section{Behavior of QIP}

This section is dedicated to evidence for good communication compilations for $\QIP$.

\subsection{Communication Upper Bounds for QMAM}

This technique of compressing the locality of the projectors by the rank of the stabilizer group can be generalized to other classes aside from $\QMA[k]$. For instance, consider the class $\QMAM$ defined in Definition \ref{def:qmam}. We have the following.
\begin{fact}(\cite{MarriottWatrous2005})\label{qipeqmam}
    $\QIP[\poly(n)](2/3, 1/3, 1) = \QMAM_{p}(1,\frac{1}{2} + 2^{-r})$ for some polynomials $r,p$.
\end{fact}

This makes $\QMAM$ an enticing class to study in order to prove that our communication upper bounds hold for general $\QIP$. To talk about this class, we first need the following lemma about representing Merlin's strategy.

\begin{lemma}\label{lem:qmam-strat}
    Merlin's strategy for $\QMAM$ protocols can be uniquely characterized by density matrices $\rho_0,\rho_1 \in \cD(\cH_1\otimes\cH_2)$ such that $\Tr_{\cH_2}(\rho_0) = \Tr_{\cH_2}(\rho_1)$, where $\cH_1\otimes\cH_2$ are the two message registers. These states $\rho_j$ represent the state Arthur receives if the coin has outcome $j$. Now let $\rho_{in} = \ketbra{x}{x} \otimes \ketbra{0^a}{0^a}$. The probability of acceptance by Arthur can thus be written as
    \begin{equation}
        \frac{1}{2}\Tr(V_0^\dagger \Pi^ZV_0(\rho_0 \otimes \rho_{in})
        + V_1^\dagger \Pi^ZV_1 (\rho_1 \otimes \rho_{in})),
    \end{equation}
    where $\Pi^Z$ is the projector $\ketbra{0}{0}\otimes\II$.
\end{lemma}

\begin{proof}
    For existence, take $\rho_0,\rho_1$ to be the states that Arthur would receive conditioned on the coin measuring outcome $0,1$, respectively. One can easily see that these states must satisfy $\Tr_{\cH_2}(\rho_0)=\Tr_{\cH_2}(\rho_1)$. For uniqueness, let $\ket{m_0},\ket{m_1}$ be purifications of $\rho_0,\rho_1$ with purification register $P$. By Uhlmann's Theorem \cite{Uhlmann1976}, there exists a unitary $U$ acting on $P,\cH_2$ such that $\ket{m_1} = U\ket{m_0}$. So, let $M_1$ be any unitary outputting $\ket{m_0}$ on input $\ket0$. Let $M_2$ apply the unitary $U$ controlled on the outcome of the coin flip. Fixing the outcome of the coin and tracing out $P$ shows that this Merlin $(M_1,M_2)$ produces $\rho_0$ and $\rho_1$.
\end{proof}

Importantly, the techniques used to prove Fact \ref{qipeqmam} need not preserve the magic of Arthur, and could drastically increase the amount of magic Arthur uses. Regardless, we show the following:

\begin{restatethm}{thm:qmam}
    Fix the polynomial $r$ from before. Given some function $f \in \QMAM_t(1,\frac{1}{2} + 2^{-r})$ with polynomial $p$ and Arthur circuit $V_0,V_1$, there are Cliffords $C_1,C_2^{(0)},C_2^{(1)}$ and $3(t+1)$-qubit density matrices $\rho_0,\rho_1$ such that the optimal strategy for Merlin is given by
    \begin{equation}
        (C_1\otimes C_2^{(j)})\qty(\rho_j \otimes \ketbra*{0^{\dim\cH_2 - 3(t+1)}}{0^{\dim\cH_2 - 3(t+1)}})(C_1\otimes C_2^{(j)})^\dagger
    \end{equation}
    based on the coin being outcome $j$. Furthermore, these Cliffords can be efficiently computed classically in $\poly(n,t)$ time.
\end{restatethm}

\begin{proof}
Now to prove Theorem \ref{thm:qmam}, we can use a very similar proof as we used in Theorem \ref{thm:qmak}, but now using the representation of Merlin's strategy described in Lemma \ref{lem:qmam-strat}.
Lemma \ref{lem:qmam-strat} tells us that the optimal success probability for the verifier $V_0,V_1$ on input $x$ is
\begin{align}
    \pi(V,x):=\max_{\substack{\rho_0,\rho_1\in\cD(\cH_1\otimes\cH_2),\\\Tr_{\cH_2}(\rho_0)=\Tr_{\cH_2}(\rho_1)}}\frac{1}{2}\Tr(V_0^\dagger\Pi^ZV_0(\rho_0 \otimes \rho_{in})
        + V_1^\dagger\Pi^ZV_1 (\rho_1 \otimes \rho_{in})),
\end{align}
for stabilizer projectors $\Pi^{S_j} = V_iP_{acc}V_i^\dagger$. Similar to $\QMA[k]$, we can again extract out $T$ gates to get
\begin{equation}
    \pi(V,x)=2^{t-1}\max_{\substack{\rho_0,\rho_1\in\cD(\cH_1\otimes\cH_2),\\\Tr_{\cH_2}(\rho_0)=\Tr_{\cH_2}(\rho_1)}}\Tr(\Pi^{S_0}\rho_0\otimes \rho_{in} \otimes \ketbra*{T^t}{T^t})+\Tr(\Pi^{S_1}\rho_1\otimes \rho_{in} \otimes \ketbra*{T^t}{T^t}),
\end{equation}
for stabilizer projectors $\Pi^{S_i}$ which have groups of rank at most $t+1$. Let $S_i^{(j)}$ be the projection of $S_i$ onto the $j$-th message from Merlin. Then, applying Lemma $\ref{lem:stab-grp-comp}$ to the group $\langle S_0^{(1)},S_1^{(1)}\rangle$, which has rank at most $2(t+1)$, and separately to $S_0^{(2)}$, $S_1^{(2)}$ gives Clifford unitaries $C_1\in\cL(\cH_1),C_2^{(0)},C_2^{(1)}\in\cL(\cH_2)$ such that
\begin{equation}
    \Pi^{S_0'}:=(C_1\otimes C_2^{(0)}\otimes\II)\Pi^{S_0}(C_1\otimes C_2^{(0)}\otimes\II)^\dagger,\qquad\Pi^{S_1'}:=(C_1\otimes C_2^{(1)}\otimes\II)\Pi^{S_1}(C_1\otimes C_2^{(1)}\otimes\II)^\dagger
\end{equation}
both act nontrivially on only at most $2(t+1)$ qubits of $\cH_1$ and $t+1$ qubits of $\cH_2$. Therefore,
\begin{equation}
    \pi(V,x)=2^{t-1}\max_{\substack{\rho_0,\rho_1\in\cD(\cH_1\otimes\cH_2),\\\Tr_{\cH_2}(\rho_0)=\Tr_{\cH_2}(\rho_1)}}\Tr(\Pi^{S_0'}\rho_0'\otimes \rho_{in} \otimes \ketbra*{T^t}{T^t})+\Tr(\Pi^{S_1'}\rho_1'\otimes \rho_{in} \otimes \ketbra*{T^t}{T^t}),
\end{equation}
where $\rho_j'=(C_1\otimes C_2^{(j)})\rho_i(C_1\otimes C_2^{(j)})^\dagger$. Like our previous result, this implies that $\rho_0',\rho_1'$ need only have support on $3(t+1)$ qubits with ancillas padded on the rest of the qubits.
\\\\
Additionally, our Cliffords are again efficiently computable using the dependence on Lemma \ref{lem:stab-grp-comp}.
\end{proof}

\subsection{Clifford QIP is in P}
Next, we prove that $\QIP_0$, the subset of $\QIP$ containing non-adaptive Clifford Arthurs, can be solved in polynomial time. In doing so, we choose useful tools to reason about how magic affects QIP. The key ingredient in this proof is showing that the optimal strategy for Merlin is achieved at a set of Clifford circuits, giving us an efficient description of a linear program we can solve using the symplectic representation of the stabilizer group.

The first part of the proof is understanding what the representations of Clifford Arthur and Merlin look like.

\begin{lemma}\label{lem:arthurrep}
    For any Arthur defined in $\QIP_0[k](c,s,o)$, the representation of his strategy, $A_{\acc}$, can be written as
    \begin{equation}A_{\acc} = 2^{m - \rank S}\Pi^S\end{equation}
    or
    \begin{equation}A_{\acc} = 0\end{equation}
    for some stabilizer projector $\Pi^S$ and $m \in \mathbb{Z}_{\ge 0}$.
\end{lemma}

\begin{proof}
    Let $C_1,...,C_k$ be Arthur's Clifford unitaries and $P_{\acc}$ be the measurement operator. From  Definition \ref{representation},
    \begin{equation}J_A(P_{\acc},A_1,...,A_k) = \Tr_{\Memory}\qty(J(\sqrt{P_{\acc}}B)J(\sqrt{P_{\acc}}B)^\dagger),\end{equation}
    where
    \begin{equation}B := (C_k \otimes \II)(\II \otimes C_{k-1} \otimes \II)...(C_1 \otimes \II)\qty(\II \otimes \ket*{0^{p(|x|)}}).\end{equation}
    From Definition \ref{choi},
    \begin{equation}J(B) = 2^{kp(|x|)/2}(B \otimes \II)\ket{\Omega}\end{equation}
    Notice that $\ket{\Omega}$ is a stabilizer state, $B$ is a Clifford isometry, and thus before the projector the state is a stabilizer state. Then, since $P_{\acc}$ is a stabilizer projector, by Fact \ref{stabstateonstabproj},
    \begin{equation}J(\sqrt{P_{acc}}B) = 2^{\frac{kp(|x|)}{2}-\frac{m}{2}}\ket{\psi}\end{equation}
    for some stabilizer state $\ket{\psi}$ and $m \in \mathbb{Z}_{\ge 0}$ or it is equal to $0$. If it is the latter, note that the proof is complete. Otherwise, using Fact \ref{traceoutstabproj}, we have
    \begin{equation}A_{acc} = 2^{kp(|x|)-m}\Tr_{\Memory}(\ketbra{\psi}{\psi}) = 2^{\rank A - m}\Pi^A\end{equation}
    for some stabilizer projector $\Pi^A$.
\end{proof}
\begin{lemma}[Clifford Uhlmann's Theorem]\label{uhlmann}
    For any two stabilizer states $\ket{\psi},\ket{\phi}$ and register $A$, if

    \begin{equation}\Tr_A(\ketbra{\psi}{\psi}) = \Tr_A(\ketbra{\phi}{\phi})\end{equation}
    then there exists a Clifford $C$, acting on $A$ such that

    \begin{equation}(C \otimes \II)\ket{\psi} = \ket{\phi}.\end{equation}
    
\end{lemma}

\begin{lemma}[Characterization of Clifford Merlin]\label{lem:merlinrep}
    For any stabilizer projector $\Pi^S$, acting on the register 
    $M_{1...k} \otimes A_{1...k-1}$, with a group $S$ which contains 
    only stabilizers $s$ that satisfy the following:
    \begin{itemize}
        \item If $s$ is non-identity on an Arthur message register 
        $A_i$, then $s$ must be non-identity within a Merlin 
        register $M_j$ such that $j > i$,
    \end{itemize}
    there exists a Merlin strategy that uses Clifford unitaries 
    $C_1,...,C_k$ s.t.
    \begin{equation}
        J_M(C_1,...,C_k) = 
        2^{|A_{1...k-1}|}
        \frac{\Pi^S}{\Tr(\Pi^S)} \otimes \II_{A_k}.
    \end{equation}
\end{lemma}

\begin{proof}
We proceed by induction over the number of rounds $k$.

For $k = 1$, take a pure stabilizer state $\ket{\psi_1}$ 
which is a purification of $\Pi^S$, such that
\begin{equation}
    \Tr_{\Memory}(\ketbra{\psi_1}{\psi_1})
    = \frac{\Pi^S}{\Tr(\Pi^S)}.
\end{equation}
Simply choose the Clifford $C_1$ that produces 
$\ket{\psi_1^*}$, so that
\begin{equation}
    J_M(C_1) = \frac{\Pi^S}{\Tr(\Pi^S)}
    \otimes \II_{A_1}.
\end{equation}
Thus the base case holds.
\\\\
For any arbitrary integer $k \ge 2$, consider $\Pi^S$ 
defined above. By Fact \ref{traceoutstabproj} and using 
that our projector must be identity on $A_{k-1}$ after 
tracing out $M_k$,
\begin{equation}
    \Tr_{M_k}\left(\frac{\Pi^S}{\Tr(\Pi^S)}\right)
    =
    \frac{\Pi^{S'}}{\Tr(\Pi^{S'})}
    \otimes \frac{\II_{A_{k-1}}}{2^{|A_{k-1}|}}
\end{equation}
for some stabilizer projector $\Pi^{S'}$ whose stabilizers 
all satisfy the condition above. This implies that
\begin{equation}
    \Tr_{M_k}\left(
        2^{|A_{1...k-1}|}\frac{\Pi^S}{\Tr(\Pi^S)}
    \right)
    =
    2^{|A_{1...k-2}|}
    \frac{\Pi^{S'}}{\Tr(\Pi^{S'})}
    \otimes \II_{A_{k-1}}.
\end{equation}
Apply the inductive hypothesis to say that this operator 
itself is the representation of a completely Clifford 
Merlin on $k-1$ rounds. Let $C_1,...,C_{k-1}$ be his 
Clifford circuits. To finish, we take that Merlin, 
extend it by an identity on the final round, and then 
prove there exists a suitable Clifford to give us our 
$k$-round representation.
\\\\
Our first observation is that this smaller Merlin can 
be written as
\begin{equation}
    \frac{\Pi^{S'}}{\Tr(\Pi^{S'})}
    =
    \Tr_{\Memory}
    (\ketbra{\psi_{k-1}^*}{\psi_{k-1}^*})
\end{equation}
where $\ket{\psi_{k-1}}$ is the normalized Choi vector 
of the smaller Merlin, which is a pure stabilizer state 
since our $B$ map is a Clifford isometry.
Now we extend our strategy by $1$ round by applying 
$\II$ on Arthur's last message and sending it back. 
Let
\begin{equation}
    \ket{\alpha} = \ket{\psi_{k-1}}
    \otimes \ket{\Phi^+}_{M_k,A_{k-1}},
\end{equation}
where $\ket{\Phi^+}_{M_k,A_{k-1}}$ consists of 
$|A_{k-1}|$ normalized EPR pairs. Our new extended 
representation can be written as
\begin{equation}
    2^{|A_{1...k-1}|}
    \Tr_{\Memory}(\ketbra{\alpha^*}{\alpha^*})
    \otimes \II_{A_k}.
\end{equation}

Next we take the normalized stabilizer state purification, 
$\ket{\psi_k}$, of $\Pi^S$ from Definition \ref{purestabproj}. 
Note that the purification register is at most as large 
as the memory Merlin has, and we can always pad it such 
that the registers are equal in size. Now we apply 
Lemma \ref{uhlmann} to say that because
\begin{equation}
    \Tr_{\Memory,M_k}(\ketbra{\psi_k}{\psi_k})
    =
    \Tr_{\Memory,M_k}(\ketbra{\alpha^*}{\alpha^*})
    =
    \frac{\Pi^{S'}}{\Tr(\Pi^{S'})}
    \otimes \frac{\II_{A_{k-1}}}{2^{|A_{k-1}|}},
\end{equation}
there exists a Clifford $C_k^*$, acting on 
$\Memory,M_k$ s.t.
\begin{equation}
    \ket{\psi_k} = (C_k^* \otimes \II)\ket{\alpha^*}.
\end{equation}
Taking the complex conjugate gives
\begin{equation}
    \ket{\psi_k^*} = (C_k \otimes \II)\ket{\alpha}.
\end{equation}

To conclude, note that defining a Merlin with 
$(C_1,...,C_k)$ and letting $B$ denote his combined 
isometry excluding the unused register $A_k$, we have
\begin{equation}
    J(B) = 2^{|A_{1...k-1}|/2}
    (C_k \otimes \II)\ket{\alpha}
    = 2^{|A_{1...k-1}|/2}\ket{\psi_k^*}.
\end{equation}
Thus
\begin{align}
    J_M(C_1,...,C_k)
    &= 2^{|A_{1...k-1}|}
    \Tr_{\Memory}(\ketbra{\psi_k}{\psi_k})
    \otimes \II_{A_k} \\
    &= 2^{|A_{1...k-1}|}
    \frac{\Pi^S}{\Tr(\Pi^S)} \otimes \II_{A_k},
\end{align}
and our proof is complete.
\end{proof}
\begin{lemma}[Characterization of Clifford Non-Measuring Arthur]\label{lem:nonmeasurearthur}
    For any stabilizer projector $\Pi^S$, acting on the register $M_{1...k} \otimes A_{1...k}$, with a group $S$ which contains only stabilizers $s$ that satisfy the following:
    \begin{itemize}
        \item If $s$ is non-identity on a Merlin message register $M_i$, then $s$ must be non-identity within a Arthur register $A_j$ such that $j > i$.
    \end{itemize}
    There exists a non-measuring Arthur strategy that uses Clifford unitaries $C_1,...,C_k$ s.t.

     \begin{equation}
        J_{\tilde A}(C_1,...,C_k) = 
        2^{|M_{1...k}|}
        \frac{\Pi^S}{\Tr(\Pi^S)}
    \end{equation}
\end{lemma}
\begin{proof}
    This proof is completely symmetric to the one above.
\end{proof}
\begin{theorem}\label{optmerlinisclifford}
    Given $f \in \QIP_0[k](c,s,o)$, for every input 
    $x \in \{0,1\}^*$, there exists an optimal strategy 
    for Merlin using Cliffords $C_1,...,C_k$.
\end{theorem}
\begin{proof}
    From \cite{Gutoski2006}, the optimal probability of success 
    in this protocol is given by
    \begin{equation}
        \min\{p \in [0,1] : A_{acc} \preceq pR, 
        \text{for some } R \in 
        \mathcal{S}_k(M_{1...k},A_{1...k})\}
    \end{equation}
    where $\preceq$ is the L\"owner partial ordering. This 
    expression can be thought of as a dual program, where 
    choices of $R$ give certificate upper bounds of the 
    optimal success probability. Because Arthur is Clifford, 
    by Lemma \ref{lem:arthurrep}, his representation is either 
    $A_{acc} = 0$ or can be written as 
    $A_{acc} = 2^{\alpha}\Pi$ for some integer 
    $\alpha$ and stabilizer projector $\Pi$.

    If $A_{acc} = 0$, then every Merlin strategy achieves 
    success probability $0$. Thus any choice of Clifford 
    unitaries $C_1,...,C_k$, such as the identity at every 
    round, is optimal, and the result holds trivially.

    Otherwise, the proof will proceed by finding a good 
    certificate $R$ and then finding a Clifford Merlin 
    that achieves this probability.
    \\\\
    Take any generators $g_1,...,g_d$ that generate the group 
    of $\Pi$. Fix an ordering of the registers as 
    $A_k,M_k,...,A_1,M_1$.
    With respect to this ordering, we will run Gaussian 
    elimination on the matrix formed with the symplectic 
    vector representations of these generators as rows. 
    We do this to find a suitable basis to observe the 
    subspaces to which Merlin and Arthur contribute. 
    After running our Gaussian elimination, each stabilizer
    row is either $\II$, in which case we can safely discard 
    it, or has a leading one on some qubit. We will call any 
    qubit that contains at least one leading one a ``pivot.'' 
    Notice that each pivot qubit is associated with either 
    one or two pivot rows. In the latter case, by taking 
    reduced row echelon form, the Pauli operators on the 
    pivot qubit in these two rows are $X$ and $Z$, up to phase.
    This is pictured in Figure \ref{fig:stabilizer-echelon}.

    \begin{figure}[t]
        \centering
        \renewcommand{\arraystretch}{1.35}
        \setlength{\arraycolsep}{12pt}
        $\displaystyle
        \begin{array}{c|c|c|c|c|c|c}
            & A_3 & M_3 & A_2 & M_2 & A_1 & M_1\\
            \hline
            g_1 & \boxed{X} & X & \II & \II & \II & \II\\
            g_2 & \boxed{Z} & Z & \II & \II & \II & \II\\
            g_3 & \II & \II & \boxed{Y} & \II & \II & \II\\
            g_4 & \II & \II & \II & \boxed{X} & X & \II\\
            g_5 & \II & \II & \II & \boxed{Z} & Z & \II\\
            g_6 & \II & \II & \II & \II & \II & \boxed{Z}\\
            \hline
            g_7 & \II & \II & \II & \II & \II & \II
        \end{array}
        $
        \caption{
        An example of the post-Gaussian elimination of the 
        matrix. The boxed entries are the Pauli operators 
        on the leading one ``pivot'' qubits.
        }
        \label{fig:stabilizer-echelon}
    \end{figure}

    Split the rows into ones with pivots in the $M$ registers 
    and ones with pivots in the $A$ registers. Call the 
    projector onto the $M$ rows $\Pi_M$ and the one onto 
    the $A$ rows $\Pi_A$. Notice that the leading non-identity 
    register of any nontrivial product of rows is the register 
    containing the leftmost pivot among those rows. Thus, 
    every non-identity stabilizer in the group of $\Pi_M$ 
    has its leading non-identity register in a Merlin 
    register, and similarly every non-identity stabilizer 
    in the group of $\Pi_A$ has its leading non-identity 
    register in an Arthur register.

    From Fact \ref{stabprojform} we can now write:
    \begin{equation}
        A_{acc} = 2^{\alpha}\Pi 
        = 2^{\alpha}\Pi_M\Pi_A.
    \end{equation}
    Consider
    \begin{equation}
        R = 2^{|M_{1...k}|}\frac{\Pi_A}{\Tr(\Pi_A)},
    \end{equation}
    which, from Lemma \ref{lem:nonmeasurearthur} and Fact 
    \ref{repiff} we know is a valid non-measuring Clifford 
    Arthur strategy, and specifically is a matrix in 
    $\mathcal{S}_k(M_{1...k},A_{1...k})$. We can choose 
    $p := 2^{\alpha-|M_{1...k}|}\cdot\Tr(\Pi_A)$. Notice that
    \begin{equation}
        \Pi_M\Pi_A \preceq \Pi_A
    \end{equation}
    since applying another commuting projector only 
    decreases (or maintains) the eigenvalues. This implies that
    \begin{equation}
        A_{acc} \preceq 
        p\,2^{|M_{1...k}|}\frac{\Pi_A}{\Tr(\Pi_A)}
        = pR.
    \end{equation}
    Thus our optimal success probability is at most 
    $2^{\alpha-|M_{1...k}|}\cdot\Tr(\Pi_A)$.

    However, by Lemma \ref{lem:merlinrep} there is a Merlin 
    using Cliffords $C_1,...,C_k$ with representation
    \begin{equation}
        J_M(C_1,...,C_k)
        = 2^{|A_{1...k}|}\frac{\Pi_M}{\Tr(\Pi_M)}.
    \end{equation}
    Since the generators of $\Pi_A$ and $\Pi_M$ are disjoint 
    subsets of an independent generating set, we have
    \begin{equation}
        \Tr(\Pi_A\Pi_M)
        =
        \frac{\Tr(\Pi_A)\Tr(\Pi_M)}
        {2^{|A_{1...k}|+|M_{1...k}|}}.
    \end{equation}
    Then the success probability of such a Merlin by 
    Fact \ref{qipsuccessprobability} is
    \begin{align}
        &\Tr\left(
            A_{acc}\,2^{|A_{1...k}|}
            \frac{\Pi_M}{\Tr(\Pi_M)}
        \right)\\
        &= \Tr\left(
            2^\alpha\Pi_M\Pi_A
            2^{|A_{1...k}|}\frac{\Pi_M}{\Tr(\Pi_M)}
        \right)\\
        &= \frac{2^{\alpha+|A_{1...k}|}
            \Tr(\Pi_A\Pi_M)}{\Tr(\Pi_M)}\\
        &= \frac{2^{\alpha+|A_{1...k}|}
            \Tr(\Pi_A)\Tr(\Pi_M)}
            {2^{|A_{1...k}|+|M_{1...k}|}\Tr(\Pi_M)}\\
        &= 2^{\alpha-|M_{1...k}|}\cdot\Tr(\Pi_A).
    \end{align}
    Thus, because there exists a Clifford Merlin strategy 
    that achieves $p$, this is exactly the optimal probability, and the result 
    holds for every input $x$.
\end{proof}
Another useful consequence of this proof is that the success probability is written as some power of $2$, thus the following holds:
\begin{corollary}\label{optsuccessis2}
    The optimal success probability for any $x\in \{0,1\}^*$ and $f \in \QIP_0[k](c,s,o)$ is always $2^{\alpha}$ for some $\alpha \in \mathbb{Z}$ or it is $0$.
\end{corollary}

We now put together our findings to conclude that this class is polynomial-time solvable.
\begin{restatethm}{thm:qipinp}
    $\QIP_0[\poly(n)](2/3,1/3,1) \subseteq \sP$
\end{restatethm}
\begin{proof}
    The proof strategy is to apply our previous lemmas and 
    transform the problem into a symplectic vector space such 
    that a polynomial-sized linear system gives us our solution.
    \\\\
    Take any function $f \in \QIP_0[\poly(n)](2/3,1/3,1)$ 
    and $x \in \{0,1\}^*$. By Theorem 
    \ref{optmerlinisclifford}, we only need to look over 
    Clifford strategies for Merlin; thus our entire circuit 
    is always a Clifford applied to the state 
    $\ket{0^{(2k+2)p(|x|) - |x|}}\ket{x}$, mapping our 
    input stabilizer state to some output stabilizer state. 
    Additionally, by Corollary \ref{optsuccessis2}, because 
    the optimal probability is either zero or a power of $2$, 
    determining if the optimal success probability is $=1$ 
    is equivalent to determining if it is $\ge \frac{2}{3}$.
    \\\\
    Our input state is stabilized by 
    $G := \langle Z_1,...,Z_{(2k+2)p(|x|) - |x|},
    \Stab(x)\rangle$ where $\Stab(x)$ is the set of 
    stabilizers for our $x$. To ensure perfect completeness, 
    we need the final state to be in the $+1$ eigenspace 
    of $Z_{\gamma}$, where $\gamma$ is Arthur's designated 
    output qubit in his private memory.
    \\\\
    We will design a linear system over our choices of 
    Merlin's Cliffords backwards from $Z_{\gamma}$ and 
    attempt to map it to some stabilizer $s \in G$. 
    This will complete our result.
    \\\\
    As a reminder, Cliffords correspond, up to Pauli signs,
    to $2n \times 2n$ symplectic matrices where $n$ is the 
    number of qubits they act on. Thus we can compute in 
    polynomial time each matrix for Arthur's Cliffords. 
    As for Merlin, since we only fix a single stabilizer, 
    we need not enforce the symplectic condition, but only 
    the invertible one. In particular, any non-identity 
    Pauli can only be conjugated to any other non-identity Pauli 
    by a Clifford.
    \\\\
    The trick to enforcing invertibility is to show that 
    Merlin has a perfect strategy, i.e., one that sends 
    some stabilizer $s$ to $Z_{\gamma}$, if and only if 
    he has a strategy that at every round on some prefix 
    of the protocol the stabilizer on his memory is 
    non-identity. Using this, the linear system can simply 
    choose any Pauli string for the message to Arthur 
    since invertibility is guaranteed.
    \\\\
    One direction of this is trivial: strategies that 
    enforce $Z$ on the first Merlin private qubit are 
    simply a subset of possibly Merlin strategies. For 
    the opposite, assume there exists Merlin Cliffords 
    $M_1,...,M_k$ that send some $s \in G$ to $Z_{\gamma}$. 
    First, notice that for some suffix of the protocol 
    Merlin might not be able to do anything. If Arthur 
    conjugates $Z_{\gamma}$ to something that is $\II$ 
    on Merlin's memory and message register, he might 
    as well apply $\II$ until his message register 
    becomes non-identity.
    \\\\
    If for the entire backward propagation Merlin has 
    $\II$ on his registers, then propagate $Z_{\gamma}$ 
    backwards through all of Arthur's Cliffords, keeping 
    track of the Pauli sign, and accept if and only if 
    the resulting signed Pauli belongs to $G$. Otherwise, 
    let $j$ be the first round encountered in the backward 
    propagation at which Merlin's message register is 
    non-identity.
    \\\\
    Let $s_1,...,s_j$ be the stabilizers at each of the 
    first $j$ rounds after being conjugated by Arthur 
    and Merlin. Notice that if we replace the restriction 
    of each of these stabilizers to Merlin's private 
    memory on the prefix before $M_j$ by $Z$ on his first 
    private qubit and identity on the remaining memory 
    qubits, leaving the message register and Arthur's 
    memory unchanged, then the restriction to Merlin's 
    registers is always non-identity. Thus, there exists 
    a Clifford mapping the restriction of each modified 
    stabilizer before Merlin's action to its restriction 
    after Merlin's action. Since Arthur does not act on 
    Merlin's private memory, these modifications do not 
    affect Arthur's conjugations.

    We can also always choose our initial $s$ to have 
    $Z$ on its first Merlin private qubit, since $Z$ is 
    in the generator set on this qubit. At round $j$, 
    Merlin can conjugate his modified non-identity Pauli 
    to the original non-identity Pauli obtained from 
    backward propagation, thereby removing the extra 
    $Z$ on his memory. Thus, we have found a Merlin 
    that always has non-identity on his register 
    throughout this prefix.
    \\\\
    Let $A_i$ be the matrix representations of Arthur's 
    Cliffords, let $z_{\gamma}$ be the vector representation 
    of our target output stabilizer, and let $g_i$ be the 
    vector representation of the $i$-th generator of our 
    input group. Order the coordinates of Arthur's 
    symplectic vector so that the first $2p(|x|)$ 
    coordinates correspond to the message register and 
    the remaining $2p(|x|)$ coordinates correspond to 
    Arthur's private memory. Let $t_i$ denote the 
    symplectic vector on these registers immediately 
    before Arthur applies $A_i$.

    We compute $j$ beforehand by running backwards through 
    our rounds. If no round is non-identity for Merlin, 
    we decide membership in $G$ as described above. 
    Otherwise, our system of linear equations can be 
    written as follows:
    \begin{equation}
        A_j^{-1}...A_k^{-1}z_{\gamma} = t_j
    \end{equation}
    \begin{equation}
        A_{i-1}^{-1}
        \begin{bmatrix}
            m_1^i \\
            \vdots \\
            m_{2p(|x|)}^i \\
            t_i^{2p(|x|)+1} \\
            \vdots \\
            t_i^{4p(|x|)}
        \end{bmatrix}
        = t_{i-1}
        \qquad \text{for } 2 \leq i \leq j
    \end{equation}
    \begin{equation}
        P_V t_1 = \sum_{\ell} \alpha_{\ell} P_V g_{\ell}
    \end{equation}
    where $m_l^i,\alpha_{\ell} \in \mathbb{F}_2$ for all 
    $i,l,\ell$, and $P_V$ denotes the projection onto 
    the symplectic coordinates of Arthur's private memory 
    register, whether applied to $t_1$ or to the global 
    stabilizer vectors $g_{\ell}$.

    Notice that we only need to enforce membership in 
    the initial stabilizer group on Arthur's private 
    memory, since Merlin's first Clifford can conjugate 
    any non-identity Pauli on his memory and message 
    register to a stabilizer of his initial all-zero 
    state. Furthermore, since Merlin acts on a non-identity 
    Pauli at round $j$, we can choose the sign of this 
    conjugation arbitrarily. Thus, any solution to the 
    linear system can be made to produce the correct 
    stabilizer sign.
    \\\\
    This is a linear system of a polynomial number of 
    variables and equations over the field $\mathbb{F}_2$, 
    and thus this algorithm runs in polynomial time, 
    completing our proof.
\end{proof}

\section{Conclusion}

In this paper, we extend our understanding of the relationship between communication and magic in different quantum verification classes. Furthermore, we show several mathematical techniques and compilation techniques to reduce the proof lengths Arthur requires from Merlin. In our first section, we show that $\QMA[2]$ and $\QMA[k]$ are indeed similar to $\QMA$ by only requiring linear in $k,t$ qubits of communication via a Clifford gadget applied to the witnesses. We prove the $\QMA[k]$ result by a different approach than the original \cite{ABNO2022} paper, which relies on the observation that the stabilizer projector is high rank. We prove the $\QMA[2]$ result by extraction by observing that extraction for $4$ parties is possible if the registers are unbalanced in size. Our second section offers a couple of pieces of evidence for the same communication bounds for $\QIP$. We first show that $\QMAM$, a simple class equal to $\QIP$, has the same compilation using compression. Lastly, we utilize the Choi representation framework of $\QIP$ to show that $\QIP$ with non-adaptive Clifford Arthur is not only solvable in polynomial time, but also that Merlin optimal strategies themselves can be Clifford.

\section{Open Questions}

Our main open question from this body of work is:

\openqitem{Can the communication in $\mathrm{QIP}$ be upper bounded by $O(kt)$?}
Our current results in this direction hold for edge cases, such as no magic, or the simpler class $\QMAM$. Trying to apply compression to general $\QIP$ by first compiling the protocol to a $\QMAM$ instance does not work, as the amount of magic in the compiled instance could increase massively. Our extraction technique relied heavily on a $4$-party observation; extraction has no apparent higher-party result. In general, extraction at more than $3$ parties is significantly harder to understand. The compression technique itself also does not seem to apply well to $\QIP$ as the rank of the projector could be much lower than for our $\QMA[k]$ and $\QMAM$ cases. Resolving this will most likely require a more powerful gadget that correlates across rounds in the protocol.

For our Choi representation techniques, we have questions about their limits for $\QIP$ and how this relates to the more general model of quantum games. 

\openqitem{Can the amount of magic Merlin needs for an optimal witness in quantum verification be upper bounded by Arthur's magic complexity?}

\openqitem{Can the same results for communication in $\QIP$ be lifted to general quantum games dependent on the total magic the two parties use?}

For our use of extraction, we are interested in higher-party extraction ($\ge 5$) following the open questions of \cite{BFG06}. A resolution to this could imply sharper bounds for $\QMA[k]$ that do not rely on the number of output qubits and could help in the study of $\QIP$.

\openqitem{Are there similar extractions guarantees for more than $4$ parties when parties are sufficiently imbalanced?}

To extend our proof of Clifford $\QIP$ belonging to $\PTIME$, we also wonder how the classical simulation of $\QIP$ scales with respect to magic complexity.

\openqitem{Can we place $\QIP$ in $\DTIME(f(n,t))$ for some known function $f$?}

We believe that progress or solutions to these questions will greatly advance our understanding of quantum information, communication, and magic complexity.

\section{AI Disclosure}

The ideas defining this paper were not found using AI, but AI was used in part for literature review, editing, and assisted with a few ideas in the proofs of our Clifford $\QIP$ result. These ideas were then distilled and simplified.

\section{Acknowledgements}

We thank Aadi Anand for helpful discussions. CN is partially supported by NSF CAREER Award IIS2541127.

\bibliographystyle{alpha}
\bibliography{references}

\end{document}